\documentclass[preprint,12pt]{elsarticle}
\usepackage{graphicx}
\usepackage{amsmath}
\usepackage{hyperref}
\usepackage{lineno}
\usepackage{amssymb}
\usepackage{cleveref}
\usepackage{bm}
\usepackage{appendix} 
\usepackage{multirow}

\begin{document}

\begin{frontmatter}

\title{Infinite Series Solution of the Time-Dependent Radiative Transfer Equation in Anisotropically Scattering Media}

\author[1]{Vladimir Allakhverdian\corref{cor1}}
\ead{allaxwerdian@yandex.ru}

\author[1]{Dmitry V. Naumov\corref{cor2}}
\ead{dnaumov@jinr.ru}

\cortext[cor1]{Corresponding author}
\cortext[cor2]{Principal corresponding author}

\address[1]{JINR, Dubna, Russia 141980}

\begin{abstract}
We solve the radiative transfer equation (RTE) in anisotropically scattering media as an infinite series. Each series term represents a distinct number of scattering events, with analytical solutions derived for zero and single scattering. Higher-order corrections are addressed through numerical calculations or approximations.

The RTE solution corresponds to Monte Carlo sampling of photon trajectories with fixed start and end points.
Validated against traditional Monte Carlo simulations, featuring random end points, our solution demonstrates enhanced efficiency for both anisotropic and isotropic scattering functions, significantly reducing computational time and resources.
The advantage of our method over Monte Carlo simulations varies with the position of interest and the asymmetry of light scattering, but it is typically orders of magnitude faster while achieving the same level of accuracy.
The exploitation of hidden symmetries further accelerates our numerical calculations, enhancing the method's overall efficiency.

In addition, we extend our analysis to the first and second moments of the photon's flux, elucidating the transition between transport and diffusive regimes.
\end{abstract}

\begin{keyword}
Radiative transfer equation \sep Green's function \sep Monte Carlo simulations \sep Anisotropic scattering media \sep Computational methods in radiative transfer
\end{keyword}

\end{frontmatter}

\section{Introduction}
\begin{figure}[!h]
\centering
\includegraphics[width=0.5\linewidth]{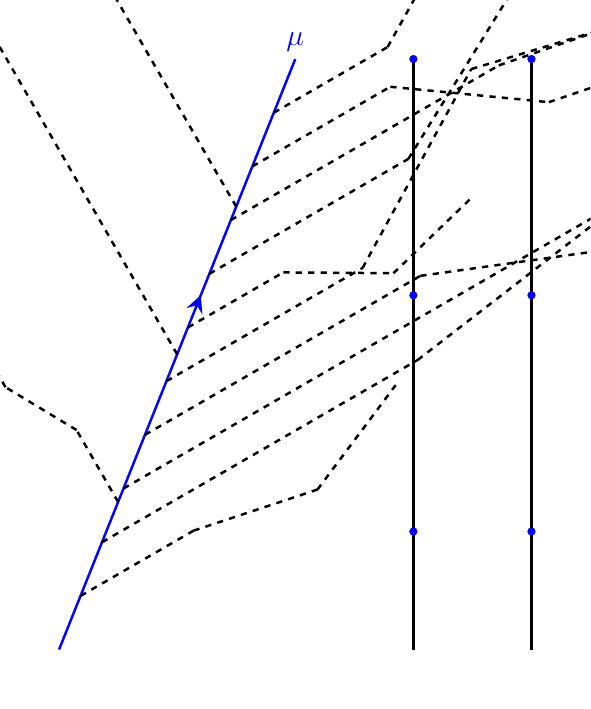}
\caption{A schematic representation of a neutrino telescope, which consists of detectors (depicted as blue circles on vertical strings) designed to capture Cherenkov photons emitted by charged particles in a medium. The figure illustrates random light trajectories with scattering events; notably, very few of these photons might hit the detectors due to their small size. }
\label{fig:nt_scheme}
\end{figure}

Light propagation in media is a subject of profound significance across multiple disciplines in physics, including astrophysics, biophotonics, particle physics, and applied research. To optimize particle detector responses, simulations of photon transport are often indispensable. In this context, Monte Carlo techniques have been a prevalent choice~\cite{doi:10.1118/1.595361,PhysRevLett.97.018104,Martelli_book,ALLISON2016186,Blyth:2021gam}.
Monte Carlo methods excel when detectors offer nearly full $4\pi$ coverage of the light source. However, their efficiency dwindles for detectors sparsely distributed within the media, as seen in the case of neutrino telescopes~\cite{PhysRevLett.113.101101,Baikal-GVD:2021zsq,KM3Net:2016zxf,ANTARES:2011hfw} and schematically shown in~\cref{fig:nt_scheme}. These scenarios necessitate extensive computational resources to attain accurate results.
To address this challenge, we propose an alternative, novel method based on solving the linear time-dependent integro-differential Radiative Transfer Equation (RTE) that governs light propagation in media.

The RTE originates from the more general nonlinear Boltzmann equation, first devised in 1872 to describe out-of-equilibrium thermodynamic systems. By making the assumption that target medium particles remain unchanged after photon scattering, the Boltzmann equation is linearized, leading to what we now know as the RTE~\cite{LernerTrigg1991}.

Despite its application across various physics domains for many years~\cite{Chandrasekhar,Case,Ishimaru,10.1145/2010324.1964951}, the RTE still lacks an analytic solution for most cases, except for the limited domain of the diffusion approximation. Consequently, numerical solutions are the standard approach.
In addition to the Monte Carlo method, various numerical methods such as finite-element~\cite{SURYAMOHAN20117364}, finite-difference~\cite{Hielscher_1998}, discrete-ordinate~\cite{GANAPOL2011693}, discontinuous finite element~\cite{Wang2023}, and lattice Boltzmann~\cite{Ymeli2023a,Feng2021,Ymeli2023b} methods are available.

However, these methods often demand extensive CPU memory due to the need for integral discretization across six variables. Even with just 100 points per variable, this results in the storage and manipulation of $10^{12}$ floating-point numbers.

Recently, solutions for the Green's function of the RTE have been derived for isotropic and unidirectional point sources in three-dimensional anisotropically scattering media~\cite{Liemert:12,PhysRevE.86.036603}. These solutions, expressed in terms of analytic functions, still require numerical integration. The method involves expanding both the solution and the scattering probability function using spherical harmonics and reducing the RTE to a matrix integrable equation. However, this method becomes particularly challenging for highly anisotropic scattering probability functions. More recently, we found the exact solution for one-dimensional time-dependent RTE~\cite{ALLAKHVERDIAN2023108726}.

In this study, we propose a novel method to solve the RTE by decomposing it into an infinite series. Each term in this series represents a distinct number of scattering events, denoted by $n\in (0,\infty)$. Remarkably, our approach mirrors the concept of Feynman diagrams used in Quantum Field Theory. Extending this parallel, we observe that the $n$-th order scattering can be conceptualized as a multi-dimensional integral, which encompasses all conceivable photon trajectories, akin to the Feynman path integrals in quantum mechanics.

Fundamentally, our method illuminates the connection between the integro-differential nature of the RTE and the multiple scatterings inherent in the Monte Carlo technique, while significantly enhancing computational efficiency. Beyond the precise computation of photon flux at any given location and time, we also offer a Monte Carlo generator for plotting all potential photon paths between predetermined start and end points.

We provide analytical expressions for the zeroth-order (no scattering) and first-order (single scattering) terms.
Remarkably, the first-order term is a singular function, featuring a delta-function. This raises concerns about the stability of numerical solutions if this singularity is disregarded, a point seemingly overlooked in existing literature to the best of our knowledge. We also delve into the efficient calculation of multi-dimensional integrals, a crucial aspect of our approach.

To validate our calculations and demonstrate the computational advantages of our method, we perform cross-checks with Monte Carlo simulations. Notably, our primary motivation for this work was to develop an efficient algorithm for light ray tracing in neutrino telescopes, where detectors are extremely sparse, and traditional Monte Carlo methods prove highly inefficient.

The structure of this article is as follows. In~\cref{sec:RTE}, we transform the integro-differential RTE into an integral equation. In~\cref{sec:expansion_series_arbitrary_source}, we present a formal solution to this equation for an arbitrary source, expressed as an infinite series with closed-form terms. In~\cref{sec:expansion_series_unidirectional}, we provide an explicit solution for a unidirectional impulse source, highlighting its physical interpretation as a series of scattering events. This section culminates in the presentation of the solution as a series of products involving the probability of no absorption, Poisson distributions, and factors dependent on all possible photon trajectories.

In~\cref{sec:moments}, we compute the first and second moments of the photon's flux by summing the entire infinite series of scattering events. We demonstrate that the realm of the transport regime is delineated by an effective inverse scattering length, reflecting the asymmetry in light scattering within the medium. As time progresses, light transport transitions to a diffusive mode, characterized by an additional effective inverse scattering length. These calculated moments offer profound insights into the transition between transport and diffusive regimes.

In~\cref{sec:regimes}, we analyze various limiting cases and regimes and give some useful approximations to the true solution. In~\cref{sec:hidden_symmetries} we discuss new hidden symmetries of the solution found and their implicitations. In~\cref{sec:numerical_results} we discuss numerical results. In particular, in~\cref{sec:comparison}, we validate our method through Monte Carlo simulations and assess its convergence rate. In~\cref{sec:quantify_orders} we study how many scatterings one has to take into account for an accurate estimate of the signal.
Finally, a discussion of obtained results and conclusions are drawn in~\cref{sec:conclusions}.

For reader's convenience, technical details are elaborated in the appendices~\labelcref{sec:transform_to_integral,app:first_order_unidirectional_derivation,%
app:single_scatter_vector_form_derivation,sec:time_integrals,%
sec:Henyey-Greenstein-properties,app:moments_HG,app:moments_n_scattering_HG,app:nested_time_integrals,%
app:moments_light_flux,app:derivation_time_integrated_flux,app:probabilistic_derivaion,%
sec:observables,sec:test_orders}.

\section{Time-dependent Radiative Transfer Equation}
\label{sec:RTE}
A solution of the light propagation problem is encoded in the six-dimensional function $L({\bm{r}},t,\hat{\bm{s}})$, defined as energy flux density at space-time point $({\bm{r}},t)$ along direction given by the unit vector $\hat{\bm{s}}$.
$L$ obeys the time-dependent RTE
\begin{equation}
\label{eqref:RTE_0}
    \left(\frac{\partial}{c \partial t}+\hat{\bm{s}}\cdot\bm{\nabla}+\mu_t\right)L({\bm{r}},t,\hat{\bm{s}}) = S({\bm{r}},t,\hat{\bm{s}}) + \mu_s \hat{V}_{\hat{\bm{s}}\hat{\bm{s}}'}L({\bm{r}},t,\hat{\bm{s}}'),
\end{equation}
where
$\hat{V}_{\hat{\bm{s}}\hat{\bm{s}}'}$ is a collision integral operator, acting on a function $L$, defined as
\begin{equation}
    \hat{V}_{\hat{\bm{s}}\hat{\bm{s}}'}L({\bm{r}},t,\hat{\bm{s}}') \equiv \int\limits_{\bm{S}^2} d\hat{\bm{s}}' f(\hat{\bm{s}}\cdot\hat{\bm{s}}')L({\bm{r}},t,\hat{\bm{s}}').
\end{equation}
$S({\bm{r}},t,\hat{\bm{s}})$ is source function, assumed to be known.
$c$ is the velocity of photons in the medium.
$\mu_t = \mu_a+\mu_s$ is the total attenuation coefficient due to absorption ($\mu_a$)  and scattering ($\mu_s$) coefficients.
$f(\hat{\bm{s}}\cdot\hat{\bm{s}}')$ is the probability density function for a photon moving in the direction specified by the unit vector  $\hat{\bm{s}}'$ to scatter into direction $\hat{\bm{s}}$.

Two other functions characterizing the light intensity are based on the definition of $L({\bm{r}},t,\hat{\bm{s}})$.
These are the fluence rate
\begin{equation}
	\label{eq:fluence_rate}
	\Phi(\bm{r},t) = \int\limits_{\bm{S}^2} d\hat{\bm{s}}\;L({\bm{r}},t,\hat{\bm{s}})
\end{equation}
and the fluence
\begin{equation}%
	\label{eq:fluence}
	F(\bm{r}) = \int\limits_{0}^\infty dt\;\Phi({\bm{r}},t).
\end{equation}
In~\cref{eq:fluence}, the lower bound of the integral, $t=0$, corresponds to the initial moment at which light begins to propagate through the medium.

The integro-differential~\cref{eqref:RTE_0} can be transformed into an integral equation (see the details in appendix~\labelcref{sec:transform_to_integral})
\begin{equation}
    \label{eqref:RTE_4}
        L(\bm{r},t,\hat{\bm{s}}) =  L_0(\bm{r},t,\hat{\bm{s}})+\Delta L(\bm{r},t,\hat{\bm{s}}),
 \end{equation}
where
\begin{equation}
\begin{aligned}
    L_0(\bm{r},t,\hat{\bm{s}})     &= c\int\limits_{0}^{t} dt'  e^{-c\mu_t(t-t')}S({\bm{r}-c\hat{\bm{s}}(t-t')},t',\hat{\bm{s}}),\\
    \Delta L(\bm{r},t,\hat{\bm{s}}) &= c\mu_s \hat{V}_{\hat{\bm{s}}\hat{\bm{s}}'}\int\limits_{0}^{t} dt' e^{-c\mu_t(t-t')}L({\bm{r}-c\hat{\bm{s}}(t-t')},t',\hat{\bm{s}}').
\end{aligned}
\label{eq:RTE_integral}
\end{equation}

\section{Iterative solution}
\label{sec:expansion_series}
\subsection{The expansion series for arbitrary source}
\label{sec:expansion_series_arbitrary_source}

In the solution of~\cref{eqref:RTE_4}, we employ an iterative method where $$L = \lim\limits_{n\to\infty} L^{(n)}$$ represents the limit of the iterative process as the number of iterations approaches infinity. Here,
\begin{equation}
    L^{(n)} = L_0 + \Delta L^{(n)}
\label{eqref:RTE_5}
\end{equation}
where $L^{(n)}$ denotes the approximation of $L$ at the $n$-th iteration and
\begin{equation}
\Delta L^{(n)}(\bm{r},t,\hat{\bm{s}}) = c\mu_s \hat{V}_{\hat{\bm{s}}\hat{\bm{s}}'}\int\limits_{0}^{t} dt' e^{-c\mu_t(t-t')}L^{(n-1)}({\bm{r}-c\hat{\bm{s}}(t-t')},t',\hat{\bm{s}}').
\label{eqref:RTE_5a}
\end{equation}
The superscript $n$ indicates the iteration level, with $n$ starting from zero and postulating $L^{(-1)}=0$.
Substituting reccursively $L^{(n-1)}$ from~\cref{eqref:RTE_5} into~\cref{eqref:RTE_5a} one can obtain that
\begin{equation}
\Delta L^{(n)} = \sum\limits_{k=1}^n\delta L^{(k)},
\label{eqref:RTE_5b}
\end{equation}
where each $\delta L^{(k)}$ term depends upon absorption and scattering medium properties.
Remarkably, $\delta L^{(k)}$ is determined by $L_0$, which could be found analytically for any source $S$ function as given by the first line in~\cref{eq:RTE_integral}. Therefore,~\cref{eqref:RTE_5,eqref:RTE_5b} provide a series approximation to the true solution of~\cref{eqref:RTE_0}. The general expression for $\delta L^{(n)}$ reads as follows
\begin{equation}
\label{eq:Ln_general}
    \delta L^{(n)}  = (c\mu_s)^ne^{-c\mu_t t}\left(\prod_{i=1}^n\int\limits_0^{t_{i+1}} dt_i\right)e^{c\mu_t t_1}
\left(\prod_{j=1}^{n}\hat{V}_{\hat{\bm{s}}_{j+1}\hat{\bm{s}}_{j}}\right)L_0(\delta\bm{r},t_1,\hat{\bm{s}}_1),
\end{equation}
where $\delta\bm{r} = \bm{r}-c\hat{\bm{s}}(t-t_n)-c\hat{\bm{s}}_n(t_n-t_{n-1})\dots -c\hat{\bm{s}}_2(t_2-t_1)$.
In~\cref{eq:Ln_general} we denoted $t=t_{n+1}$ and $\bm{s}=\bm{s}_{n+1}$.  One should expect $\lim\limits_{n\to\infty}\delta L^{(n)}=0$ if the series converges.

Let us write explicitly first four orders to gain some insight.
The zero order reads
\begin{equation}
    \label{eqref:zero_order_solution}
    L^{(0)} = L_0.
\end{equation}
The first order correction reads
\begin{equation}
    \label{eqref:first_order_solution}
    \begin{aligned}
    \delta L^{(1)}& = c\mu_s\int\limits_{0}^{t} dt_1 e^{-c\mu_t (t-t_1)} \hat{V}_{\hat{\bm{s}}\hat{\bm{s}}_1}
 L_0(\delta\bm{r},t_1,\hat{\bm{s}}_1),
    \end{aligned}
\end{equation}
where $\delta\bm{r} = {\bm{r}-c\hat{\bm{s}}(t-t_1)}$.
The second order correction reads
\begin{equation}
    \label{eqref:second_order_solution}
    \begin{aligned}
    \delta L^{(2)}  & = (c\mu_s)^2 \int\limits_{0}^{t} dt_2\int\limits_{0}^{t_2} dt_1 e^{-c\mu_t (t-t_1)}\hat{V}_{\hat{\bm{s}}\hat{\bm{s}}_2}\hat{V}_{\hat{\bm{s}}_2\hat{\bm{s}}_1}
L_0(\delta\bm{r},t_1,\hat{\bm{s}}_1),
   \end{aligned}
\end{equation}
where $\delta\bm{r}={\bm{r}-c\hat{\bm{s}}(t-t_2)-c\hat{\bm{s}}_2(t_2-t_1)}$.
The third order is, respectively
\begin{equation}
    \label{eqref:third_order_solution}
    \delta L^{(3)}
    = (c\mu_s)^3 \int\limits_{0}^{t} \!dt_3\!\int\limits_{0}^{t_3}\! dt_2\!\int\limits_{0}^{t_2}\! dt_1 e^{-c\mu_t (t-t_1)}
\hat{V}_{\hat{\bm{s}}\hat{\bm{s}}_3}\hat{V}_{\hat{\bm{s}}_3\hat{\bm{s}}_2}\hat{V}_{\hat{\bm{s}}_2\hat{\bm{s}}_1}
L_0(\delta\bm{r},t_1,\hat{\bm{s}}_1),
\end{equation}
where $\delta\bm{r} = \bm{r}-c\hat{\bm{s}}(t-t_3)-c\hat{\bm{s}}_3(t_3-t_2)-c\hat{\bm{s}}_2(t_2-t_1)$.

\subsection{The Green's function for unidirectional source}
\label{sec:expansion_series_unidirectional}
Let us consider the point-like source function
\begin{equation}
\label{sec:source_delta}
    S(\bm{r},t,\hat{\bm{s}}) = \delta^3(\bm{r})\delta(t)\delta^2(\hat{\bm{s}}-\hat{\bm{s}}_0)
\end{equation}
which corresponds to a photon produced at point $\bm{r}=0$ along a direction given by unit vector $\hat{\bm{s}}_0$ at time $t=0$. Other initial conditions can be easily accounted for replacing $\bm{r}\to \bm{r}-\bm{r}_0$ and $t\to t-t_0$ for any $\bm{r}_0$ and $t_0$. We display further results in the vector form. When appropriate we use an explicit coordinate system defined by $\hat{\bm{s}}_0 =(0,0,1)$, where $\bm{r}=(x,y,z)$. Then,
\begin{equation}
  \hat{\bm{s}}=\hat{\bm{s}}(\theta,\varphi) = (\sin\theta\cos\varphi,\sin\theta\sin\varphi,\cos\theta).
\end{equation}

\subsubsection{Warming Up: The First Four Scattering Terms}
\begin{figure}[!h]
\centering
\includegraphics{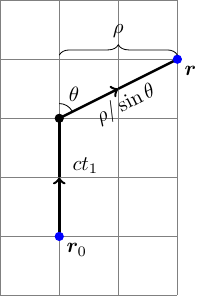}
\caption{The diagram depicts a photon's trajectory from the source point $\bm{r}_0$ to the observation point $\bm{r}$ through a single scattering event. The distance $\rho$ represents the lateral displacement in the scattering plane, and $\theta$ denotes the scattering angle. The single scattering time $t_1$ corresponds to the moment when light emitted along the initial direction $\hat{\bm{s}}_0$ changes its direction to $\hat{\bm{s}}$ towards point $\bm{r}$.}
\label{fig:single_scattering_path}
\end{figure}

Substituting~\cref{sec:source_delta} into~\cref{eqref:RTE_5} one gets a corresponding $n$-th order correction to the solution $L$. Explicitly, zero order reads
\begin{equation}
\begin{aligned}
    L_0 & = c e^{-c\mu_t t}\delta^3(\bm{r}-c\hat{\bm{s}}_0 t)\delta^2(\hat{\bm{s}}-\hat{\bm{s}}_0)\\
    & =c e^{-c\mu_t t}\frac{\delta(r-ct)}{r^2}\delta^2(\bm{\hat{r}}-\hat{\bm{s}}_0)\delta^2(\hat{\bm{s}}-\hat{\bm{s}}_0),
\end{aligned}
\label{eq:zero_order_unidirectional}
\end{equation}
which is known as the ballistic trajectory without scatterings and where $\bm{\hat{r}}=\bm{r}/|\bm{r}|$. The first order correction, assuming light emitted along $z$ axis, is found to be (interested reader may follow the derivation in~\cref{app:first_order_unidirectional_derivation}):
\begin{equation}
\begin{aligned}
\delta L^{(1)} &=c^2\mu_s e^{-c\mu_t t}f(\hat{\bm{s}}\cdot\hat{\bm{s}}_0)\int\limits_0^t dt_1\delta^3(\delta\bm{r})\\
&= c\mu_s e^{-c\mu_t t}\frac{f(\hat{\bm{s}}\cdot\hat{\bm{s}}_0)}{\rho (1-\cos\theta)}\delta(\rho-(ct-z)\cot{\frac{\theta}{2}})\delta(\varphi-\varphi^*)\\
&\times H(ct-z)H(\frac{z}{ct}-\cos\theta),
\end{aligned}
\label{eq:first_order_unidirectional}
\end{equation}
where in the first line $\delta\bm{r}=\bm{r}-c\hat{\bm{s}}_0t_1-c\hat{\bm{s}}(t-t_1)$ and in the next lines
\begin{equation}
\begin{aligned}
\rho &=\sqrt{x^2+y^2}, \quad \varphi^* = \arctan{(y/x)},
\end{aligned}
\label{eq:single_scattering_point}
\end{equation}
and $H(x)$ is the Heaviside step-function. The product of two Heaviside step-functions in~\cref{eq:first_order_unidirectional} excludes superluminal light transport. Let us note that $\delta L^{(1)}$ is proportional to a product of two delta-functions, implying that only four out of the six variables are independent for single scattering events. This observation has a straightforward geometric explanation as depicted in~\cref{fig:single_scattering_path}.
Two vectors, $\hat{\bm{s}}_0$ and $\bm{r}$, lie in a plane, and single scattering occurs exclusively within this plane. We shall refer to this plane as {\em the scattering plane}. There exists a unique moment in time, denoted as $t_1$, when light emitted along $\hat{\bm{s}}_0$ reaches point $\bm{r}$
traveling in the direction of $\hat{\bm{s}}$:
\begin{equation}
t_1 = \frac{z-\rho\cot{\theta}}{c}.
\end{equation}
Hence, when we specify a particular direction $\hat{\bm{s}}$ at point $\bm{r}$, the arrival time $t$ becomes uniquely determined and is not independent anymore:
\begin{equation}
t = \frac{z+\rho\tan{(\theta/2)}}{c}.
\end{equation}
Additionally, because the light trajectory lies within the scattering plane, the azimuth angle is also fixed.
While calculations in the explicit coordinate system can aid in visualization, it is often more convenient to express them in vector form. In this form, the first-order correction, denoted as $\delta L^{(1)}$ , can be expressed as follows:
\begin{equation}
\begin{aligned}
\delta L^{(1)} &=2c\mu_s e^{-c\mu_t t}f(\hat{\bm{s}}\cdot\hat{\bm{s}}_0)\delta^2(\hat{\bm{s}} - \hat{\bm{s}}^*)\frac{1}{|\bm{r}-c\hat{\bm{s}}_0t|^2}\\
&\times H\left((ct)^2 - \bm{r}^2\right)H\left(ct - \bm{r}\cdot\hat{\bm{s}}_0\right),
\end{aligned}
\label{eq:single_scatter_vector_form}
\end{equation}
where
\begin{equation}
\begin{aligned}
&\hat{\bm{s}}^* = \hat{\bm{s}}_0+2\frac{\bm{r} - c\hat{\bm{s}}_0t}{|\bm{r} - c\hat{\bm{s}}_0t|^2}(ct-\bm{r}\cdot\hat{\bm{s}}_0).
\end{aligned}
\label{eq:magic_direction}
\end{equation}
Times $t_1$ and $t$ are found to be:
\begin{equation}
\begin{aligned}
t_1 &= \frac{1}{2c}\left(\frac{\bm{r}\cdot(\hat{\bm{s}}_0-\hat{\bm{s}})}{1-\hat{\bm{s}}\cdot\hat{\bm{s}}_0}
+\frac{\bm{r}\cdot(\hat{\bm{s}}_0+\hat{\bm{s}})}{1+\hat{\bm{s}}\cdot\hat{\bm{s}}_0}\right),\\
t   &= \frac{1}{c}\frac{\bm{r}\cdot(\hat{\bm{s}}_0-\hat{\bm{s}})}{1-\hat{\bm{s}}\cdot\hat{\bm{s}}_0}.
\end{aligned}
\label{eq:time_single_scattering}
\end{equation}
The derivation of~\cref{eq:single_scatter_vector_form,eq:magic_direction,eq:time_single_scattering} can found in~\cref{app:single_scatter_vector_form_derivation}.
Before we conclude with a general solution, let's examine the next two orders. The second-order correction is expressed as follows:
\begin{equation}
\begin{aligned}
    \delta L^{(2)}  &
    =c^3\mu_s^2e^{-c\mu_t t}\int\limits_{0}^{t} dt_2\int\limits_{0}^{t_2} dt_1 \int\limits_{\bm{S}^2} d\hat{\bm{s}}_1 f({\hat{\bm{s}}\cdot\hat{\bm{s}}_1})f({\hat{\bm{s}}_1\cdot\hat{\bm{s}}_0})\delta^3(\delta\bm{r}),
\end{aligned}
\label{eq:second_order_unidirectional}
\end{equation}
where $\delta \bm{r} =\bm{r}-c\hat{\bm{s}}(t-t_2)-c\hat{\bm{s}}_1(t_2-t_1)-c\hat{\bm{s}}_0t_1$.
The third-order correction is given by:
\begin{equation}
\!\!\delta L^{(3)} = c^4\mu_s^3e^{-c\mu_tt}\!\!\int\limits_{0}^{t}\!\!dt_3\!\!\int\limits_{0}^{t_3}\!\!dt_2\!\int\limits_{0}^{t_2}\!\!dt_1\!\!\int\limits_{\bm{S}^2}\!\! d\hat{\bm{s}}_1\!\!\int\limits_{\bm{S}^2}d\hat{\bm{s}}_2f({\hat{\bm{s}}\cdot\hat{\bm{s}}_2})f({\hat{\bm{s}}_2\cdot\hat{\bm{s}}_1})f({\hat{\bm{s}}_1\cdot\hat{\bm{s}}_0})\delta^3(\delta\bm{r}),
\label{eq:third_order_unidirectional}
\end{equation}
where $\delta\bm{r}=\bm{r}-c\hat{\bm{s}}(t-t_3)-c\hat{\bm{s}}_2(t_3-t_2)-c\hat{\bm{s}}_1(t_2-t_1)-c\hat{\bm{s}}_0t_1$.

For an arbitrary $n$-th order:
\begin{equation}
\label{eq:deltan_0}
\begin{aligned}
\delta L^{(n)}  &= (\mu_sc)^n e^{-c\mu_t t}\left(\prod_{i=1}^n\int\limits_0^{t_{i+1}} dt_i\right)
\left(\prod_{j=1}^{n-1}\int\limits_{\bm{S}^2}d\hat{\bm{s}}_{j} f(\hat{\bm{s}}_{j+1}\cdot\hat{\bm{s}}_{j})\right) f(\hat{\bm{s}}_1\cdot\hat{\bm{s}}_{0})\times\\
&\times c\delta^3(\bm{r} - c\sum_{i=0}^{n}\hat{\bm{s}}_i(t_{i+1}-t_i)),
\end{aligned}
\end{equation}
where $\hat{s}_{n} \equiv\hat{s}$ and $t_{n+1} \equiv t$.

\subsubsection{Qualitative Analysis}
To get a further insight let us discuss the obtained expressions in~\cref{eq:zero_order_unidirectional,eq:first_order_unidirectional,%
eq:second_order_unidirectional,eq:third_order_unidirectional} in greater details.

\begin{figure}[!h]
\begin{center}
\includegraphics[width=\linewidth]{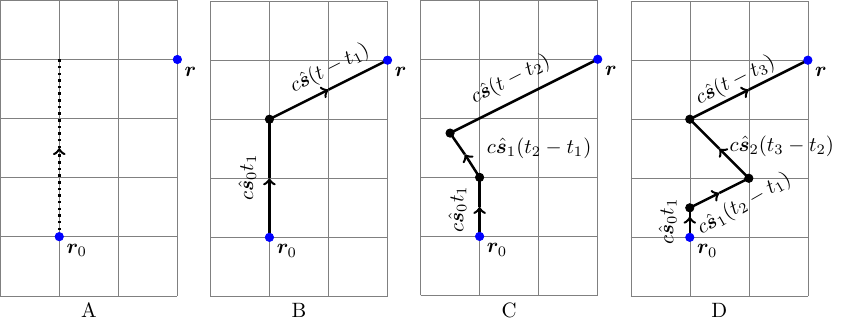}
\caption{Four photon trajectory examples from $\bm{r}_0$ to $\bm{r}$ at time $t$ with initial direction $\hat{\bm{s}}_0$:
(A) Zero scattering: $\bm{r}=\bm{r}_0+c\hat{\bm{s}}(t-t_0)$ and $\hat{\bm{s}}=\hat{\bm{s}}_0$.
(B) Single scattering at $t_1$: $\bm{r}=c\hat{\bm{s}}_0t_1+c\hat{\bm{s}}(t-t_1)$.
(C) Two scatterings: (1) at $t_1$ to $\hat{\bm{s}}_1$, (2) at $t_2$ to $\hat{\bm{s}}$,
reaching $\bm{r}=c\hat{\bm{s}}(t-t_2)+c\hat{\bm{s}}_1(t_2-t_1)$.
(D) Three scatterings: (1) at $t_1$ to $\hat{\bm{s}}_1$, (2) at $t_2$ to $\hat{\bm{s}}_2$,
(3) at $t_3$ to $\hat{\bm{s}}$,
reaching $\bm{r}=c\hat{\bm{s}}(t-t_3)+c\hat{\bm{s}}_2(t_3-t_2)+c\hat{\bm{s}}_1(t_2-t_1)$.
}
\label{fig:series_illustration}
\end{center}
\end{figure}

(i) The zero-order $L^{(0)}$ corresponds to a so-called ballistic component of the solution describing  photon propagation along the straight line trajectory $\bm{r}(t)=c\hat{\bm{s}}_0 t$  with unchanged direction $\hat{\bm{s}}=\hat{\bm{s}}_0$. The corresponding fluence and fluence rate are determined by both absorption and scattering lengths, since no scattering and no absorption should occur for zero-order
\begin{equation}
\begin{aligned}
 \Phi^{(0)}(\bm{r},t)  &= c e^{-c\mu_t t}\frac{\delta(r-ct)}{r^2}\delta^2(\bm{\hat{r}}-\hat{\bm{s}}_0),\\
	F^{(0)}(\bm{r}) &= \frac{e^{-\mu_t r}}{r^2}\delta^2(\bm{\hat{r}}-\hat{\bm{s}}_0),
  \end{aligned}
  \label{eq:fluence_rate_zero_order}
\end{equation}
where $\bm{\hat{r}}$ is the unit vector along $\bm{r}$. One could find an illustration of these considerations in~\cref{fig:series_illustration}A.

(ii) The first-order correction $\delta L^{(1)}$ corresponds to the single scattering contribution.
The scattering occurs at a particular moment in time $t_1$ when the initial direction $\hat{\bm{s}}_0$ changes to $\hat{\bm{s}}$, and the photon travels during time $t-t_1$ to reach the point $\bm{r}$ with no further scattering and absorption. Let us note once again that for given $\bm{r}$ and  $\hat{\bm{s}}$, time $t$ is not a free variable for the single scattering, as given by~\cref{eq:time_single_scattering}.
This trajectory is encoded in the argument of the $\delta$-function yielding $\bm{r}=c\hat{\bm{s}}_0t_1-c\hat{\bm{s}}(t-t_1)$.
The single scattering contribution is illustrated in~\cref{fig:series_illustration}B.
A simple but important observation about the single scattering is that it occurs in the scattering plane determined by $\hat{\bm{s}}_0$ and $\bm{r}$.
The first-order correction to the fluence rate $\delta\Phi^{(1)}\equiv \int\limits_{\bm{S}^2}d\hat{\bm{s}}\delta L^{(1)}$ reads
\begin{equation}
	\begin{aligned}
	\delta\Phi^{(1)} &= c\mu_s \frac{e^{-c\mu_t t}f(\hat{\bm{s}}^*\cdot\hat{\bm{s}}_0)\sin\theta^*}{\rho(ct-z)}H(ct-z)H(\frac{z}{ct}-\cos\theta^*)\\
                   &=2c\mu_s \frac{e^{-c\mu_t t}f(\hat{\bm{s}^*}\cdot\hat{\bm{s}}_0)}{|\bm{r}-c\hat{\bm{s}}_0t|^2}H((ct)^2 - \bm{r}^2)H(ct - (\bm{r}\cdot\hat{\bm{s}}_0)),
	\end{aligned}
  \label{eq:fluence_rate_first_order}
\end{equation}
where $\rho$ and $\varphi^*$ are determined by~\cref{eq:single_scattering_point} and
\begin{equation}
\theta^* =2\arctan{\frac{ct-z}{\rho}}.
\end{equation}

(iii) The second-order $\delta L^{(2)}$ is due to double scattering contributions.
The latter includes all allowed photon's trajectories of the following kind.
The photon with the initial direction $\hat{\bm{s}}_0$ at some moment in time $t_1$ changes its direction to an arbitrary $\hat{\bm{s}}_1$ with the corresponding probability  $f({\hat{\bm{s}}_1,\hat{\bm{s}}_0})d\hat{\bm{s}}_1$.
Then, after the time interval $t_2-t_1$ it scatters again, now into $\hat{\bm{s}}$ direction, with the corresponding probability density $f({\hat{\bm{s}},\hat{\bm{s}}_1})$, and travels to $\bm{r}$ without further scattering and absorption.
This trajectory is ensured by the argument of the $\delta$-function yielding $\bm{r}=c\hat{\bm{s}}(t-t_2)+c\hat{\bm{s}}_1(t_2-t_1)+c\hat{\bm{s}}_0t_1$ and displayed in~\cref{fig:series_illustration}C.

(iv) Finally, a trajectory with three scatterings is depicted in~\cref{fig:series_illustration}D.
The argument of the $\delta$-function ensures the trajectory $${\bm{r}=c\hat{\bm{s}}(t-t_3)+c\hat{\bm{s}}_2(t_3-t_2)+c\hat{\bm{s}}_1(t_2-t_1)}+c\hat{\bm{s}}_0t_1.$$
Although we illustrate two and three scatterings in the same plane for clarity, scatterings with $n > 2$ generally deviate from the scattering plane.

All other orders follow the same pattern and the solution reads
\begin{equation}
\label{eq:Ln_general_unidir}
    L = L_0 + \sum_{n=1}^\infty \delta L^{(n)}.
\end{equation}
\subsubsection{The Solution}
The expression in~\cref{eq:deltan_0} can be rendered more transparent by introducing dimensionless variables in place of time variables, defined as follows:
\begin{equation}
\label{eq:xi_variable}
\xi_k = 1 - \frac{t_k}{t_{k+1}}, \quad \xi_k \in (0,1).
\end{equation}
With these variables, we can rewrite it as a product of three factors (a detailed proof can be found in appendix \labelcref{sec:time_integrals}):
\begin{equation}
\delta L^{(n)} = e^{-\mu_a ct} \cdot P_n(\mu_s ct) \cdot \delta L_s^{(n)},
\label{eq:deltaLn}
\end{equation}
where

(i) $e^{-\mu_a ct}$ represents the probability of no absorption during the time interval $t$.

(ii) $P_n(\mu_s ct)$ corresponds to the Poisson distribution, which models the occurrence of $n$ random, statistically independent scatterings during the time interval $t$ with a mean rate of $c\mu_s$.

(iii) The last factor is defined as:
\begin{equation}
\label{eq:deltaLs}
\begin{aligned}
\delta L_s^{(n)} & = n! \prod_{i=1}^n \left(\int\limits_0^{1} d\xi_i(1-\xi_i)^{i-1}\right)\prod_{j=1}^{n-1}\left(\int\limits_{\bm{S}^2}d\hat{\bm{s}}_{j} f(\hat{\bm{s}}_{j+1}\cdot\hat{\bm{s}}_{j})\right) f(\hat{\bm{s}}_1\cdot\hat{\bm{s}}_{0})\times\\
&\times c\delta^3(\bm{r} - ct\bm{s}^{(n)}).
\end{aligned}
\end{equation}
where
\begin{equation}
\label{eq:s_n}
\begin{aligned}
  \bm{s}^{(n)} &=   \sum\limits_{k=0}^{n}\hat{\bm{s}}_{n-k}\xi_{n-k}\left(\prod\limits_{i=1}^{k}(1-\xi_{n-i+1})\right)\\
  &=\hat{\bm{s}}_{n}\xi_{n}+\hat{\bm{s}}_{n-1}\xi_{n-1}(1-\xi_n)+\ldots\\
  &+\hat{\bm{s}}_{1}\xi_1 (1-\xi_n)\ldots(1-\xi_{2})\\
  &+\hat{\bm{s}}_{0}\xi_0 (1-\xi_n)\ldots(1-\xi_{1})\\
\end{aligned}
\end{equation}
depends on all possible photon's trajectories with $n$ scatterings during $t$. Let us note, that $\xi_0=1$ for $t_0=0$ in agreement with~\cref{eq:xi_variable}. Notably,
\begin{equation}
    \prod_{i=1}^n \left(\int\limits_0^{1} d\xi_i(1-\xi_i)^{i-1}\right) = \frac{1}{n!},
\end{equation}
which warrants the factor $n!$ in the front of these integrals in~\cref{eq:deltaLs}. The presence of angular integrals with $\delta^3(\bm{r} - ct\bm{s}^{(n)})$ function shapes the values of these integrals.

The complete solution for the Green's function of unidirectional light source in homogeneous medium reads as follows
\begin{equation}
\label{eq:L_final_solution}
        L  = e^{-\mu_t ct}c \delta^3(\bm{r}-c\hat{\bm{s}}_0 t)\delta^2(\hat{\bm{s}}-\hat{\bm{s}}_0)
         +e^{-\mu_a ct}\sum_{n=1}^\infty P_n(\mu_s ct) \delta L_s^{(n)},
\end{equation}
where the first term denotes the unscattered light, exponentially attenuated by the total attenuation coefficient $\mu_t$, and the second term includes summation over $n\in(1,\infty)$ scatterings determined by the Poisson probability mass function $P_n(\lambda) = \frac{\lambda^n}{n!}e^{-\lambda}$ with $\lambda = \mu_s ct$ and the closed-form term $\delta L_s^{(n)}$ given in~\cref{eq:deltaLs} accounts for scattering dynamics, determined by $f(\hat{\bm{s}}'\cdot\hat{\bm{s}})$. We assume the convergence of the series as presented in~\cref{eq:L_final_solution}.
\Cref{eq:L_final_solution} is a major result of this work.
In general, calculation of integrals for $n\ge 2$ in~\cref{eq:deltaLs} has to be done numerically.
A derivation of~\cref{eq:L_final_solution} based on purely probabilistic ground can be found in~\cref{app:probabilistic_derivaion}.

\section{Moments of the Light Flux}
\label{sec:moments}
The solution outlined in~\cref{eq:L_final_solution} enables the examination of various moments of the light flux. This comprehensive analysis facilitates the study of the time evolution of mean light flux direction ($\hat{\bm{s}}$), position ($\bm{r}$), and velocity ($\bm{v}$), along with the dispersions and correlations of these observables.

Remarkably, the derived results are entirely analytical, providing clear delineation of three distinct light transport regimes: linear, diffusive, and intermediate. These regimes are elucidated through our theoretical framework, offering profound insights into the dynamics of light propagation.

\subsection{First Moments}
We define the first moment of an observable $f=\bm{r}, \bm{s}, \bm{v}$ as follows:
\begin{equation}
    \langle f(t) \rangle = \frac{1}{N} \int\limits_{\mathbb{S}^2} d\hat{\bm{s}} \int d^3\bm{r} f(\bm{r}, t, \hat{\bm{s}}) L(\bm{r}, t, \hat{\bm{s}}),
    \label{eq:first_moment_definition}
\end{equation}
where $L(\bm{r}, t, \hat{\bm{s}})$ represents the photon's flux magnitude and
\begin{equation}
    N = \int\limits_{\mathbb{S}^2} d\hat{\bm{s}} \int d^3\bm{r} L(\bm{r}, t, \hat{\bm{s}}) = ce^{-\mu_a ct}
    \label{eq:first_moment_normalization_definition}
\end{equation}
is determined using~\cref{eq:L_final_solution}.

For the first moment of the position vector, we find (please refer to~\cref{app:moments_light_flux_mean_coordinate} for details of the derivation)
\begin{equation}
    \langle\bm{r}(t)\rangle = \frac{1}{\mu_s'}(1 - \exp(-\mu_s'ct))\hat{\bm{s}}_0,
\end{equation}
where
\begin{equation}
    \mu_s' = \mu_s(1 - g)
    \label{eq:mus_effective}
\end{equation}
is the effective inverse scattering length.

Several observations can be made regarding the average coordinate $\langle\bm{r}(t)\rangle$. (i) It never exceeds one effective scattering length given by~\cref{eq:mus_effective}. This indicates a fundamental scale beyond which the photon's propagation is significantly affected by scattering. (ii) It is directly proportional to the initial direction $\hat{\bm{s}}_0$ and is null in any plane orthogonal to this initial direction. (iii) For early times, characterized by $\mu_s'ct \ll 1$, $\langle\bm{r}(t)\rangle$ is proportional to $ct$, which is consistent with the expectations from linear transport theory where scattering events are rare. Conversely, at longer times, as scattering becomes predominant, the diffusion approximation takes precedence, leading to a stabilization of the average coordinate, indicative of a transition from ballistic to diffusive light transport.

For the first moment of direction, we obtain (please refer to~\cref{app:moments_light_flux_mean_direction} for details of the derivation)
\begin{equation}
    \langle\hat{\bm{s}}(t)\rangle = \exp(-\mu_s'ct)\hat{\bm{s}}_0,
\end{equation}
and for the velocity, we find
\begin{equation}
    \langle\bm{v}(t)\rangle = c \exp(-\mu_s'ct)\hat{\bm{s}}_0.
\end{equation}
Similarly to the average coordinate, in the transport approximation, the photon's velocity matches the speed of light in the medium. Yet, transitioning to the diffusion approximation, the average velocity of the photon diminishes to zero. This signifies the photon's entry into the diffusion regime, where all directions are equivalent, and initial directional information is lost.

\subsection{Second moments}
The second moments are defined similarly to~\cref{eq:first_moment_definition} with $f=\hat{\bm{s}}^i \hat{\bm{s}}^j, \bm{r}^i\bm{r}^j$ and $\hat{\bm{s}}^i\bm{r}^j$. To avoid cluttering we omit indication of their functional time dependence.
Let us begin with the angular correlation function. We found it to be given by
\begin{equation}
    \langle \hat{\bm{s}}^i \hat{\bm{s}}^j \rangle = \frac{1}{3}(1-\exp(-\mu_s''ct))\delta_{ij} + \exp(-\mu_s''ct)\hat{\bm{s}}_0^i\hat{\bm{s}}_0^j,
\end{equation}
where $\delta_{ij}$ is the Kronicker delta symbol and
\begin{equation}
    \mu_s'' = \mu_s(1-g^2)
\end{equation}
is the second effective inverse scattering length distinct from~\cref{eq:mus_effective}. An interested reader may refer to~\cref{app:second_moments_light_flux_direction} for full details of the derivation.

At early times within the transport regime, the directional correlations align with the initial orientation, fulfilling $\langle \hat{\bm{s}}^i \hat{\bm{s}}^j \rangle = \hat{\bm{s}}_0^i\hat{\bm{s}}_0^j$. With increasing photon propagation time, a transition occurs towards the diffusion regime, rendering $\langle \hat{\bm{s}}^i \hat{\bm{s}}^j \rangle$ isotropic and equivalent to $\frac{1}{3}\delta_{ij}$. Notably, the transition timescale is governed by the parameter $\mu_s''$, diverging from $\mu_s'$.

The directional dispersion is articulated as
\begin{equation}
D(s) = \text{Tr}(\langle \hat{\bm{s}}_i \hat{\bm{s}}_j \rangle - \langle \hat{\bm{s}}_i\rangle\langle\hat{\bm{s}}_j \rangle)  = 1- e^{-2\mu_s'ct}.
\end{equation}

Regarding spatial correlations (please, refer to~\cref{app:second_moments_light_flux_space} for full details of the derivation):
\begin{equation}
    \begin{aligned}
        \langle \bm{r}^i \bm{r}^j\rangle = & \frac{2}{(\mu_s')^2(1+g)g}\times \\
        &\left[- \frac{1}{3} \left(g(g+2+\mu_sct(g^2-1)) -(1+g)^2e^{-\mu_s'ct} + e^{-\mu_s''ct}\right)\delta_{ij}\right. \\
        &\left.+ \left(g-(1+g)e^{-\mu_s'ct} + e^{-\mu_s''ct}\right)\hat{\bm{s}}_0^i\hat{\bm{s}}_0^j\right].
    \end{aligned}
\end{equation}
A care should be taking considering three limiting cases $g=\pm 1, 0$:
\begin{equation}
    \begin{aligned}
        g = 1: & \quad \langle \bm{r}_i \bm{r}_j \rangle = (ct)^2 \hat{\bm{s}}_{0i} \hat{\bm{s}}_{0j}, \\
       g = -1: & \quad \langle \bm{r}_i \bm{r}_j \rangle = -\frac{1}{2\mu_s^2}\left(1 - 2\mu_s ct - e^{-2\mu_s ct}\right) \hat{\bm{s}}_{0i} \hat{\bm{s}}_{0j}, \\
        g = 0: & \quad \langle \bm{r}_i \bm{r}_j \rangle = -\frac{2}{3\mu_s^2} \left(2 - \mu_s ct - (2 + \mu_s ct)e^{-\mu_s ct}\right) \delta_{ij} \\
               & \quad \qquad\qquad + \frac{2}{\mu_s^2} \left(1 - e^{-\mu_s ct}(1 + \mu_s ct)\right) \hat{\bm{s}}_{0i} \hat{\bm{s}}_{0j}.
    \end{aligned}
\end{equation}
For any $g$ in the transport regime:
\begin{equation}
\langle \bm{r}^i \bm{r}^j\rangle  = (ct)^2\hat{\bm{s}}_0^i\hat{\bm{s}}_0^j.
\end{equation}
Conversely, under the diffusion paradigm:
\begin{equation}
\begin{aligned}
g \ne \pm 1: & \quad \langle \bm{r}_i \bm{r}_j\rangle = \frac{2ct}{3\mu_s'}\delta_{ij} + \frac{2}{(\mu_s')^2(1+g)}(\hat{\bm{s}}_0)_i(\hat{\bm{s}}_0)_j,\\
g=-1: & \quad \langle \bm{r}_i\bm{r}_j\rangle = \frac{ct}{\mu_s}\hat{\bm{s}}_0^i\hat{\bm{s}}_0^j.
\end{aligned}
\end{equation}
Please, note that for $g = 1$, only the transport regime exists.

Furthermore, the spatial dispersion is delineated as:
\begin{equation}
    D(r) = \text{Tr}(\langle \bm{r}_i\bm{r}_j\rangle - \langle \bm{r}_i\rangle \langle\bm{r}_j\rangle),
\end{equation}
and the relative spatial dispersion quantified by
\begin{equation}
    \delta_r = \sqrt{D(r)}/\langle r\rangle.
\end{equation}
\begin{figure}[!h]
\centering
\includegraphics[width=0.8\linewidth]{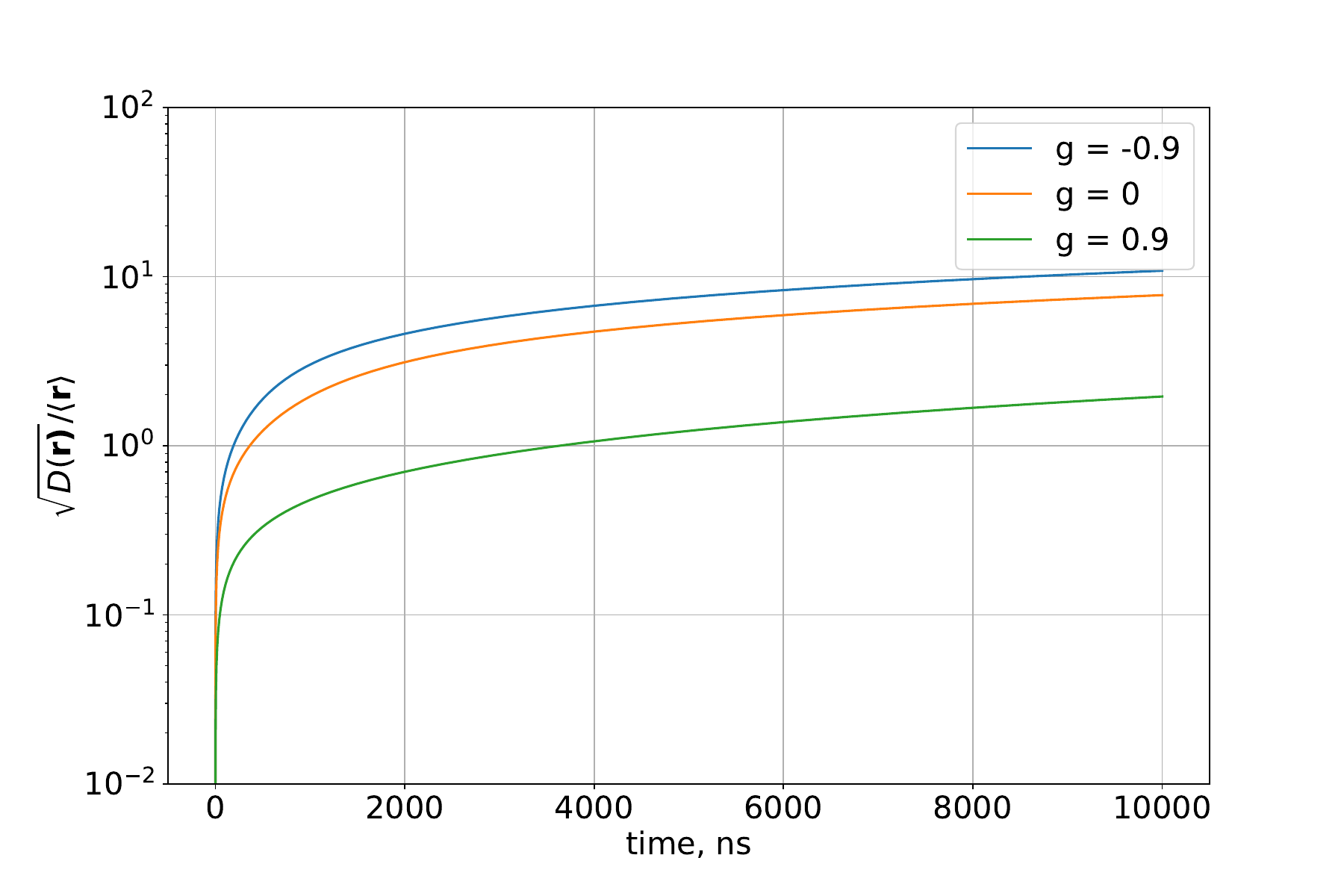}
\caption{The relative spatial dispersion $\delta_r= \sqrt{D(r)}/\langle r\rangle$ vs time for three examples of the scatteting assymetry $g=-0.9,0,0.9$.}
\label{fig:av_pos_fluc}
\end{figure}
In~\cref{fig:av_pos_fluc} $\delta_r$ is displayed as function of time for three examples of the scattering assymetry $g=-0.9,0,0.9$. One can notice that for $g = 0.9$, $\delta_r$ ascend more slowly than for other $g$ values. This slower growth is rationalized by the photon's preferential forward movement, reducing the spatial variance in comparison to the isotropic or backward scattering scenarios defined by $g = 0$ and $g = -0.9$, respectively.

Finally, we calculate the average of $\hat{\bm{s}}^i\bm{r}^j$ (please, refer to~\cref{app:second_moments_light_flux_direction_space} for full details of the derivation)
\begin{equation}
\begin{aligned}
    \langle \hat{\bm{s}}^i\bm{r}^j  \rangle =& \frac{1}{\mu_s'g}\frac{\delta_{ij}}{3}(e^{-\mu_s''ct}+g-(1+g)e^{-\mu_s'ct}\\& + \frac{1}{\mu_s'g}(e^{-\mu_s'ct} - e^{-\mu_s''ct})\hat{\bm{s}}_0^i\hat{\bm{s}}_0^j
    \end{aligned}
\end{equation}
and of their scalar product:
\begin{equation}
    \langle \bm{r}\cdot \hat{\bm{s}} \rangle =\frac{1}{\mu_s'}(1-e^{-\mu_s'ct}) = \langle \bm{r} \rangle \hat{\bm{s}}_0.
\end{equation}
Let us summarize the major outcomes of moments' calculations. (i) The first moments elucidate that the transport regime is characterized by the effective inverse scattering length, denoted as $\mu_s' = \mu_s(1 - g)$. This parameter is pivotal in defining the scale and behavior of photon transport in this regime. (ii) The second moments provide insights into both the transport and diffuse regimes, as well as delineating the transition between these two states. These moments are influenced by two distinctive effective scattering lengths, $\mu_s'$ and $\mu_s'' = \mu_s(1 - g^2)$, which is a new result.

\section{Analyzing Regimes: Limiting Cases and Approximations}
\label{sec:regimes}
\subsection{Fully Anisotropic Scattering ($g=1$) Case}
It is instructive to derive the solution in the case of fully anisotropic scattering $f(\hat{\bm{s}}\cdot\hat{\bm{s}}')=\delta^2(\hat{\bm{s}}-\hat{\bm{s}}')$, which corresponds to $g\to 1$ in the Henyey-Greenstein phase function~\cite{henyey1941diffuse} (see~\cref{eq:hg_convolution_3} in appendix~\labelcref{sec:Henyey-Greenstein-properties}).

In this case
\begin{equation}
\delta L_s^{(n)} = \delta^3(\bm{r} - c\bm{s}_0 t) \delta^2(\bm{s} - \bm{s}_0)
\end{equation}
and accounting for $\sum\limits_{n=1}^\infty P_n(\lambda) = 1 - P_0(\lambda)$, we arrive at
\begin{equation}
\label{eqref:RTE_no_scattering_delta}
L(\bm{r}, t, \hat{\bm{s}}) = c e^{-c\mu_a t} \delta^3(\bm{r} - c\hat{\bm{s}}_0 t) \delta^2(\hat{\bm{s}} - \hat{\bm{s}}_0)
\end{equation}
which notably does not depend on the scattering coefficient $\mu_s$.

This result is an immediate solution for the RTE in~\cref{eqref:RTE_0}, assuming no scattering.

\subsection{Spherical Fluence: Time-Integrated Photon Flux within a Sphere}
In this section, we focus on the calculation of the time-integrated photon flux passing through a sphere with a radius of $r$, which we refer to as the ``spherical fluence''. Mathematically, this can be expressed as (please, refer to~\cref{app:derivation_time_integrated_flux} for details of the derivation):
\begin{equation}
\begin{aligned}
G(r, \bm{s}) &\equiv \int_0^\infty dt \int_{\bm{S}^2} d\hat{\bm{r}} L(\bm{r}, t, \hat{\bm{s}}) = \frac{1}{r^2} e^{-\mu_t r} \delta^2(\hat{\bm{s}} - \hat{\bm{s}}_0) \\
&+ \frac{1}{r^2} \sum_{n=1}^\infty \left(\prod_{i=1}^n \int_0^1 d\xi_i (1-\xi_i)^{i-1}\right) \left(\prod_{j=1}^{n-1} \int_{\bm{S}^2} d\hat{\bm{s}}_j f_g(\hat{\bm{s}}_{j+1} \cdot \hat{\bm{s}}_j)\right) \\
&\times f_g(\hat{\bm{s}}_1 \cdot \hat{\bm{s}}_0) \left(\frac{\mu_s r}{|\bm{s}^{(n)}|}\right)^n \frac{e^{-\mu_t r/|\bm{s}^{(n)}|}}{|\bm{s}^{(n)}|}
\end{aligned}
\label{eq:spherical_fluence1}
\end{equation}

It's important to note that~\cref{eq:spherical_fluence1} provides an exact result, which necessitates numerical methods for its computation.

However, in the limit where $r$ significantly exceeds $\mu_s^{-1}$, we can make a simplifying approximation by setting $|\bm{s}^{(n)}|$ to its minimal value of one, and due to the convolution properties of the Henyey-Greenstein scattering phase function as demonstrated in appendix~\labelcref{sec:Henyey-Greenstein-properties}, all other integrations can be carried out exactly. This leads to the following approximation:
\begin{equation}
G(r,\bm{s}) \approx \frac{1}{r^2}e^{-\mu_t r}\left(\delta^2(\hat{\bm{s}}-\hat{\bm{s}}_0)+\sum\limits_{n=1}^\infty  \frac{(\mu_s r)^n}{n!}f_{g^n}(\hat{\bm{s}}\cdot\hat{\bm{s}}_{0})\right).
\label{eq:spherical_fluence2}
\end{equation}
The approximation described in~\cref{eq:spherical_fluence2} reveals that as we consider higher-order scatterings, they increasingly contribute to a more isotropic distribution of light.
\subsection{Extended Source}
The obtained results for the Green's function can be easily applied for an arbitrary extended source  $S(\bm{r},t,\bm{s})$ by standard technique:
\begin{equation}
L(\bm{r},t,\bm{s}) = \int d\bm{r}_0 dt_0 d\hat{\bm{s}}_0 G(\bm{r}-\bm{r}_0,t-t_0,\hat{\bm{s}}-\hat{\bm{s}}_0)S(\bm{r}_0,t,\hat{\bm{s}}),
\label{eq:flux_extended_source1}
\end{equation}
where $G$ is the Green's function.

Assuming instantaneous and unidirectional source function $S(\bm{r},t,\bm{s})=S(\bm{r})\cdot \delta^2(\hat{\bm{s}}-\hat{\bm{s}}_0)\delta(t)$ and substituting~\cref{eq:L_final_solution} for  $G$ one gets:
\begin{equation}
L(\bm{r},t,\bm{s}) =  e^{-\mu_t ct }S(\bm{r}-ct\hat{\bm{s}})\delta^2(\hat{\bm{s}}-\hat{\bm{s}}_0)+  e^{-\mu_a ct}\sum_{n=1}^\infty P_n(\mu_s ct) \delta \widetilde{L}_s^{(n)},
\label{eq:flux_extended_source2}
\end{equation}
where
\begin{equation}
\label{eq:deltaLs_extended1}
\begin{aligned}
\delta \widetilde{L}_s^{(n)} & = n! \prod_{i=1}^n \left(\int\limits_0^{1} d\xi_i(1-\xi_i)^{i-1}\right)\prod_{j=1}^{n-1}\left(\int\limits_{\bm{S}^2}d\hat{\bm{s}}_{j} f(\hat{\bm{s}}_{j+1}\cdot\hat{\bm{s}}_{j})\right) f(\hat{\bm{s}}_1\cdot\hat{\bm{s}}_{0})\times\\
&\times cS(\bm{r} - ct\bm{s}^{(n)}).
\end{aligned}
\end{equation}
For a particular case of a slow dependence of $S$ on ${\bm{r}}$: $S(\bm{r})=S$,~\cref{eq:deltaLs_extended1} can be drastically simplified because all integrals can be carried out, yielding:
\begin{equation}
L(\bm{r},t,\bm{s}) =  S\left(e^{-\mu_t ct }\delta^2(\hat{\bm{s}}-\hat{\bm{s}}_0)+  e^{-\mu_a ct}\sum_{n=1}^\infty P_n(\mu_s ct) f_{g^n}(\hat{\bm{s}}\cdot\hat{\bm{s}}_{0})\right),
\label{eq:flux_extended_source2b}
\end{equation}

\section{Hidden Symmetries}
\label{sec:hidden_symmetries}
The solution found in~\cref{eq:L_final_solution} possesses several {\em hidden} symmetries that are crucial for further accelerating numerical calculations. Let us summarize them by expressing the position vector as $\bm{r}=r\hat{\bm{r}}$, where $\hat{\bm{r}}$ is the unit vector along $\bm{r}$. One can prove that the $n$-th order contribution $\delta L^{(n)}$ transforms in a particular way under {\em scaling of the distance} $r\to \alpha r$, where $\alpha>0$ is an arbitrary scaling parameter. The transformation of $\delta L^{(n)}$ reads as follows:
\begin{equation}
    \delta L^{(n)}(\alpha r \hat{\bm{r}}, t, \hat{\bm{s}}) = \exp{\left[-\mu_t ct\left(\frac{\alpha-1}{\alpha} \right)\right]}\alpha^{n-3}\delta L^{(n)}(r \hat{\bm{r}}, \frac{t}{\alpha}, \hat{\bm{s}}).
    \label{eq:scaling_distance_transform}
\end{equation}

Since each order transforms according to its own law, this symmetry is not a general symmetry for the entire solution, as each component of the solution transforms differently. However, because we find our solution as a series of consecutive approximations, this symmetry provides a major numerical improvement. Indeed, our experience with numerical calculations of multidimensional integrals shows that one systematically has to increase the number of function calls when considering more distant points. At the same time, for shorter distances, the evaluation of integrals is typically faster. Therefore, by calculating $\delta L^{(n)}$ at some distance $r$, one immediately obtains the light flux at any other distance using~\cref{eq:scaling_distance_transform}.

Similarly, we can derive transformation formulas for changing the wavelength $\lambda_1\to\lambda_2$:
\begin{equation}
\begin{aligned}
    \delta L^{(n)}(r\hat{\bm{r}} , t, \hat{\bm{s}}|\lambda_2) = &
    \exp{\left[\left(\mu_t(\lambda_2) - \mu_t(\lambda_1)\right)c(\lambda_2)t\right]}\times\\
    &\times\frac{n(\lambda_2)}{n(\lambda_1)}\left(\frac{\mu_s(\lambda_2)n(\lambda_1)}{\mu_s(\lambda_1)n(\lambda_2)}\right)^n\delta L^{(n)}(r \hat{\bm{r}}, \frac{n(\lambda_1)}{n(\lambda_2)}t, \hat{\bm{s}}|\lambda_1),
\end{aligned}
\end{equation}
where $n(\lambda)$ is the refractive index of the medium and $\mu_s(\lambda)$ represents the dependence of the inverse scattering length on the wavelength.

There is also an evident exact symmetry for the entire scattering series under the action of a rotation matrix ${R}$:
\begin{equation}
    L(r {R}\hat{\bm{r}}, t, \hat{\bm{s}}| \hat{\bm{s}}_0) = L(r\hat{\bm{r}}, t, {R}^{-1}\hat{\bm{s}}| {R}^{-1}\hat{\bm{s}}_0).
\end{equation}

\section{Numerical Results}
\label{sec:numerical_results}
\subsection{Performance tests}
\label{sec:comparison}

To illustrate the performance, accuracy, and domain of applicability of our proposed calculation method, we estimate the expected signal for a spherical detector with realistic absorption and scattering lengths appropriate for Baikal water. Mathematical details of the calculation are summarized in appendix~\ref{sec:observables}. In this section, we consider the following working examples:

(i) We place a unidirectional instantaneous light source $\bm{s}_0=(0,0,1)$ at the origin $(0,0,0)$ m.

(ii) The center of a spherical photo-detector with a radius of $R=21$ cm is placed in four different test positions: near and far points in the forward direction, $\bm{r} = (3,0,3)$ m and $\bm{r} =(3,0,100)$ m, a side point $\bm{r} =(3,0,0)$ m, and a backward position $\bm{r} =(0,0,-3)$ m.

(iii) We choose water absorption and scattering lengths $\mu_a^{-1} = 20.9$ m and $\mu_s^{-1} = 69.26$ m, appropriate for light with a wavelength of $\lambda = 488$ nm propagating in Baikal water as an example.

(iv) The corresponding refractive index is $n=1.366$.

(v) We use the Henyey-Greenstein phase function (see~\cref{eq:Henyey-Greenstein}) with the anisotropy parameter $g=0.9$, which is close to Baikal water's property.

The multidimensional numerical integrations are performed using the VEGAS Monte-Carlo integrator~\cite{Lepage:1977sw,Lepage:2020tgj}.

As a cross-check, we employ a Monte-Carlo ray-tracing routine written in NumPy, which should converge to the true solution with infinitely large statistics.
\begin{figure}[!h]
    \centering
    \begin{tabular}{cc}
    \includegraphics[width=0.5\textwidth]{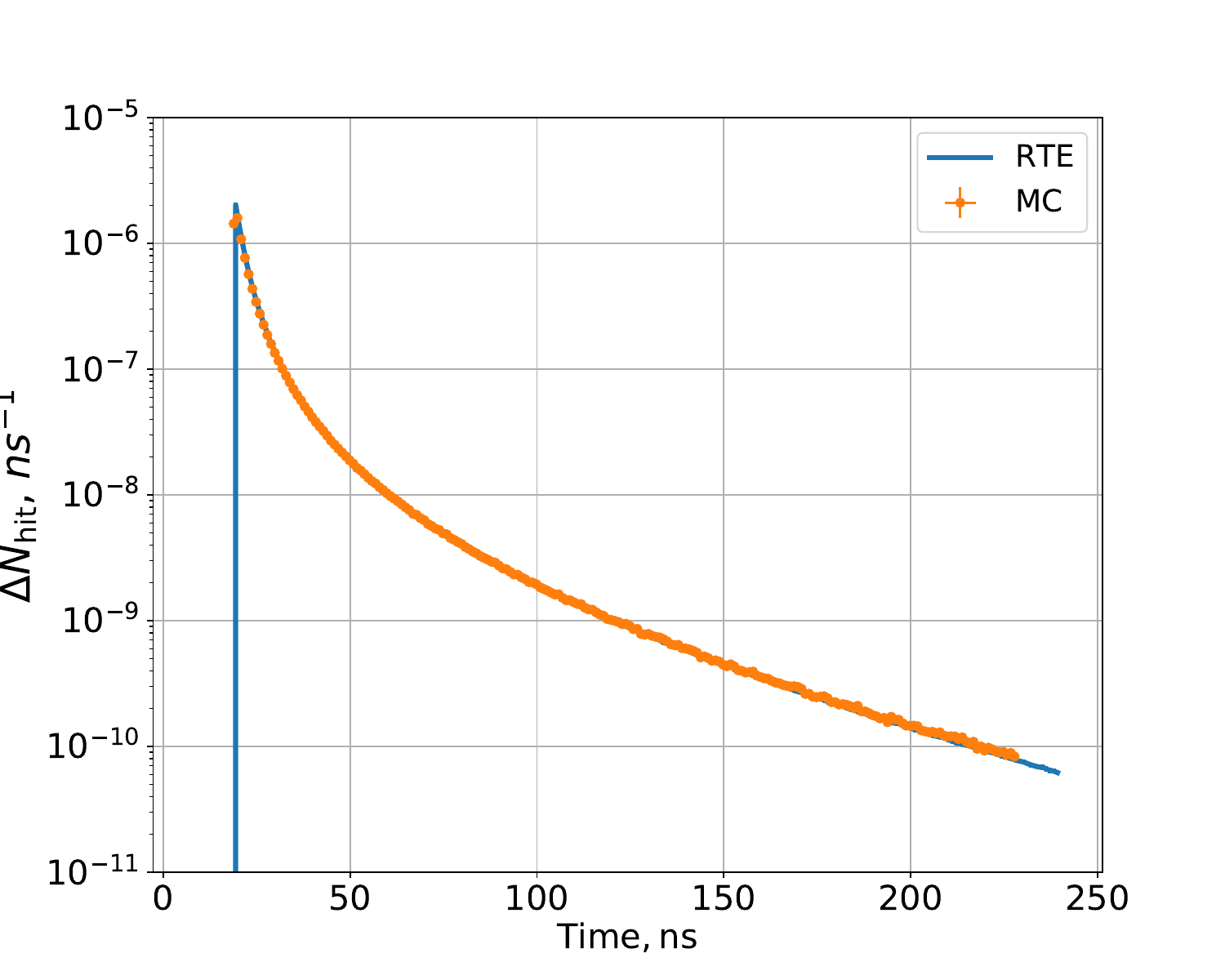}&
    \includegraphics[width=0.5\textwidth]{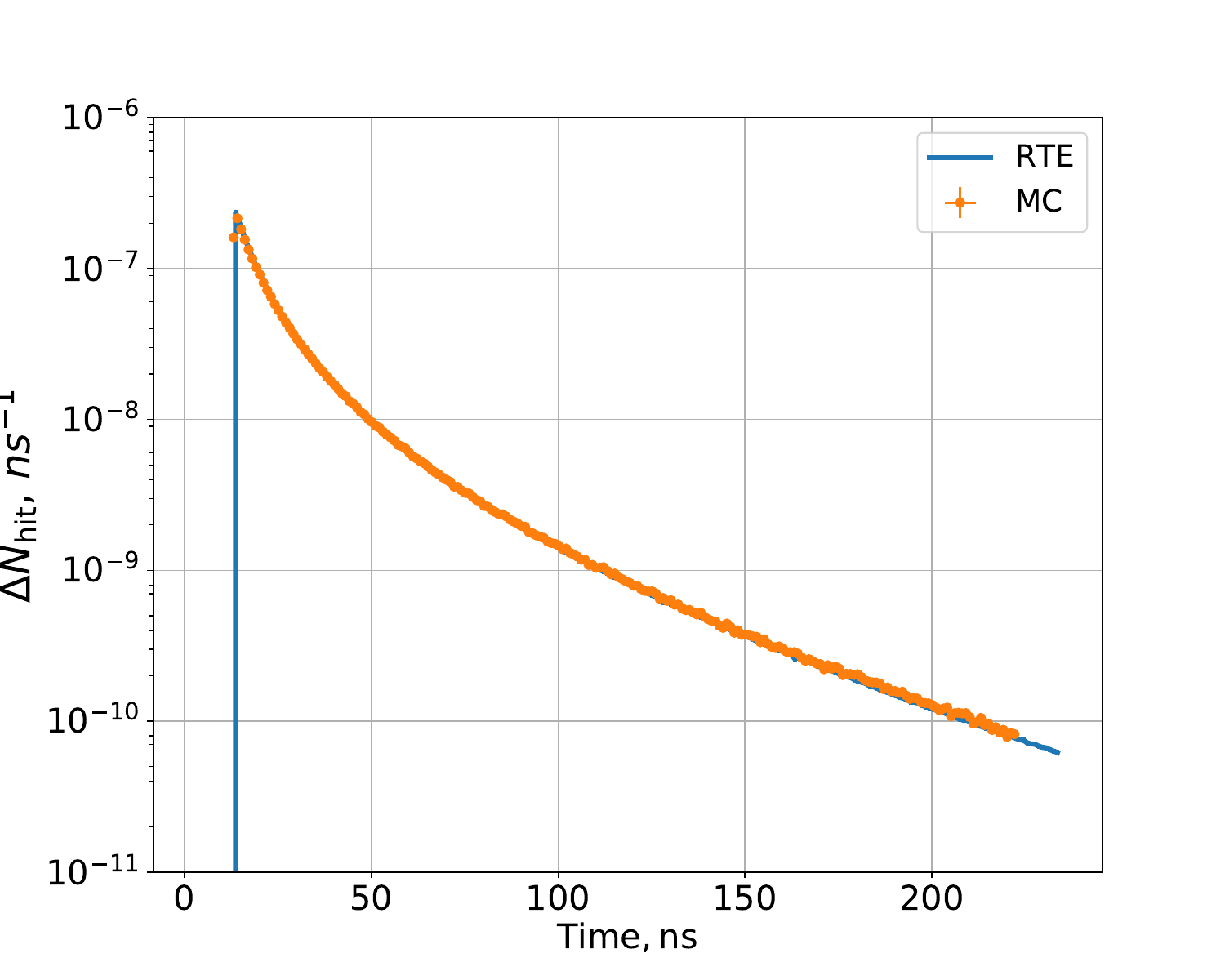}\\
    \includegraphics[width=0.5\textwidth]{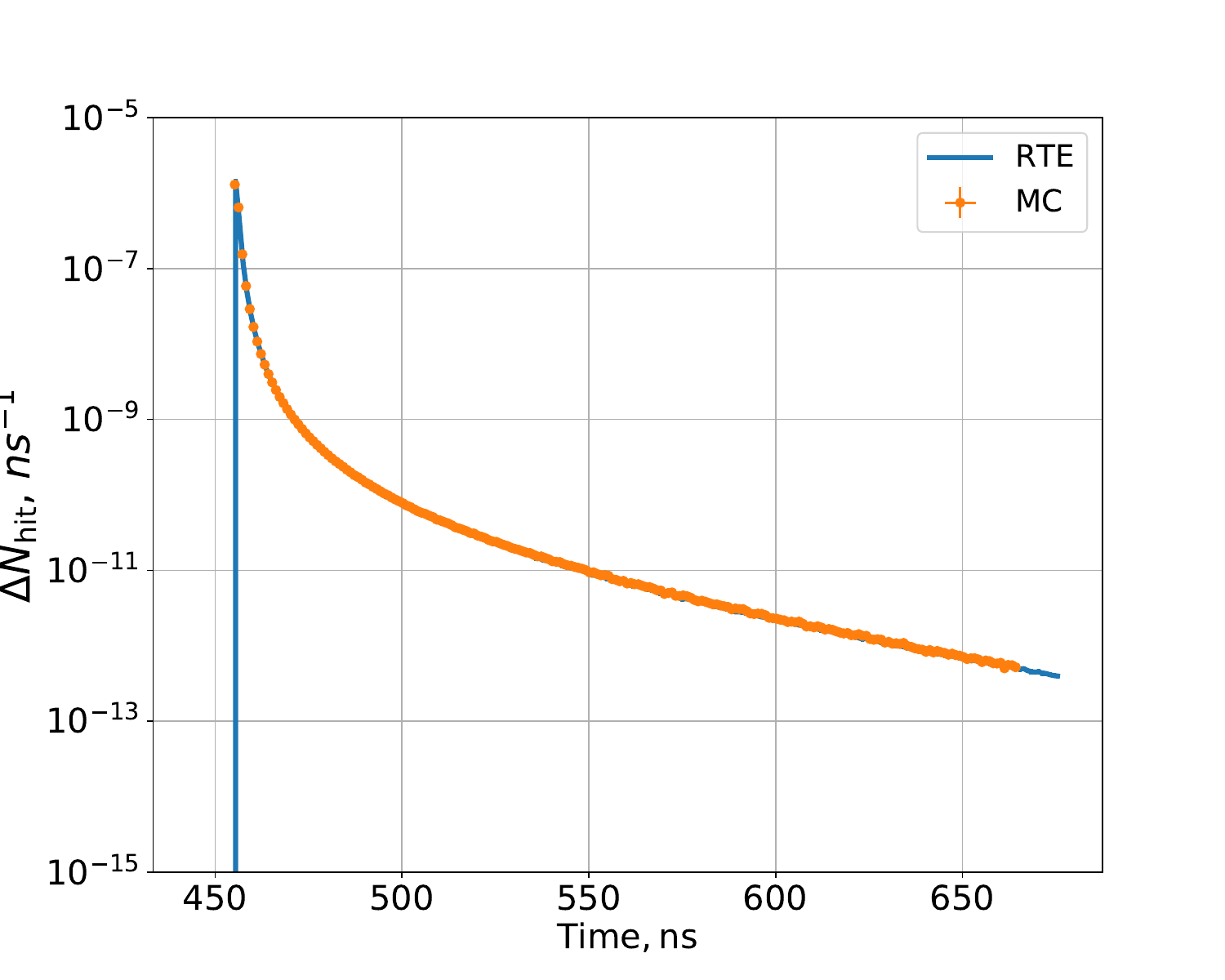}&
    \includegraphics[width=0.5\textwidth]{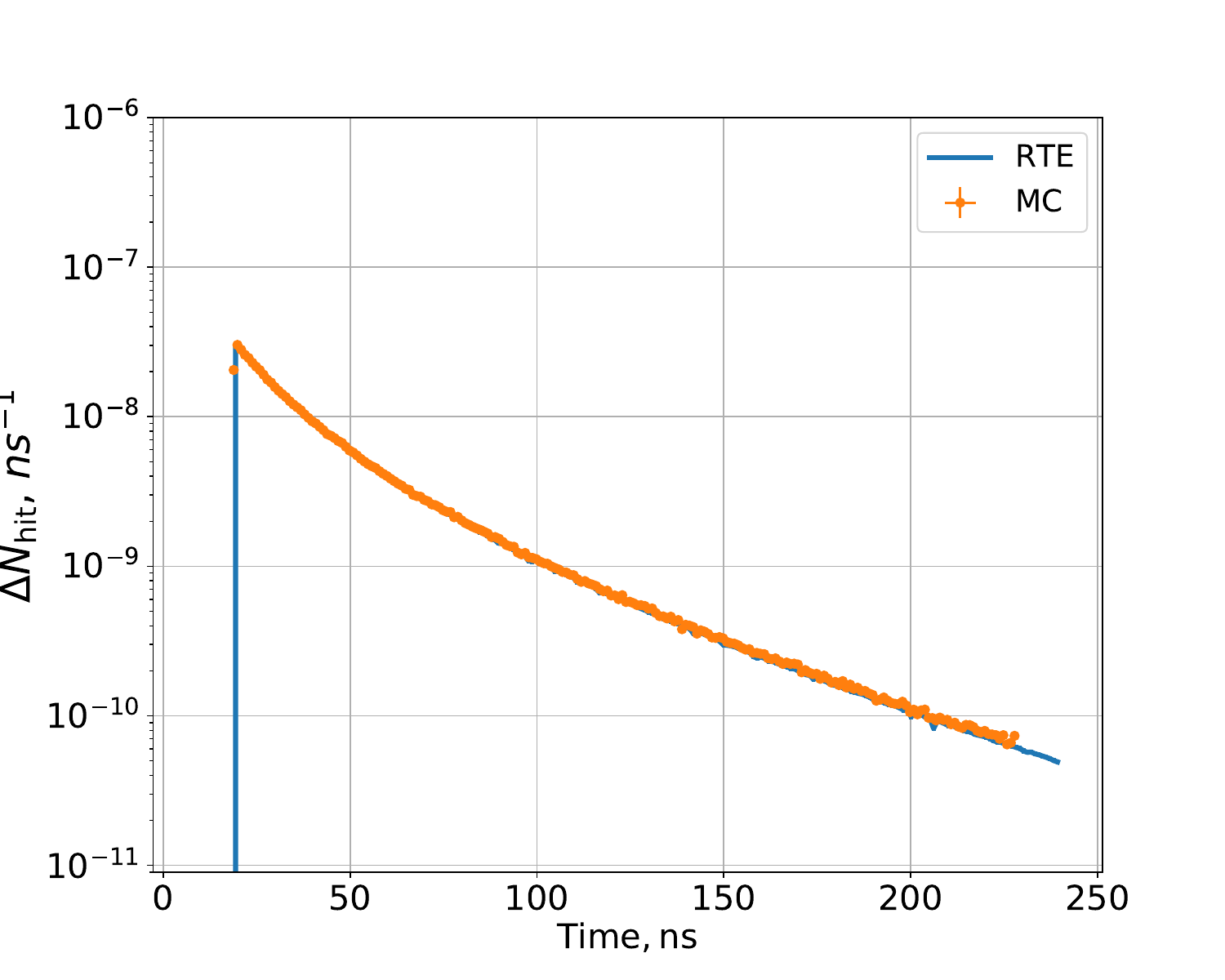}
    \end{tabular}
    \caption{Expected number of detected photons $\Delta N_\text{hit}$ integrated in one ns time bins obtained by our solution of the radiative transfer equation (blue solid line labeled as RTE) and by a Monte-Carlo ray-tracing method (brown points with error bars labeled as MC) for four points: $\bm{r} = (3,0,3)$ m (top left), $\bm{r} = (3,0,0)$ m (top right), $\bm{r} = (3,0,100)$ m (bottom left), $\bm{r} = (3,0,-3)$ m (bottom right). Note: The presence of noise in these distributions is a result of the Monte Carlo method used for VEGAS multi-dimensional integral calculations. The total calculation time for a single plot is about two hours using the RTE method, compared to 63 days with the Monte Carlo method.}
    \label{fig:RTE_MC_comparison}
\end{figure}
In~\cref{fig:RTE_MC_comparison}, we present the estimates of the expected number of detected photons, denoted as $\Delta N_\text{hit}$, integrated over one ns time bins. These estimates are obtained using our approach to solve RTE as described in~\cref{eq:DeltaN}, accounting up to four scattering orders. They are compared with results obtained from a Monte-Carlo ray-tracing method (MC).

\begin{table}[!h]
\centering
\begin{tabular}{|c|c|c|c|c|c|}
\hline
\multirow{2}{*}{Test Point $\bm{r}$, in $m$} & \multicolumn{4}{c|}{RTE} & \multirow{2}{*}{MC} \\
\cline{2-5}
                     & 2nd               & 3rd     & 4th  & Total &  \\ \hline
$(3, 0, 3)$          & $8\cdot 10^{-3}$  & $0.48$  & $0.48$  & $1.0$   & $66$ \\ \hline
$(3, 0, 0)$          & $2.6\cdot 10^{-2}$  & $1.4$  & $1.4$  & $3.0$   & $1012$ \\ \hline
$(3, 0, -3)$         & $9.8\cdot 10^{-2}$  & $5.6$  & $5.6$  & $12.1$   & $8975$ \\ \hline
$(3, 0, 100)$        & $5\cdot 10^{-2}$ & $5.7$ & $5.7$ & $13.9$ & $15$ \\ \hline
\end{tabular}
\caption{
CPU time required for RTE and MC ray-tracing calculations, normalized to the total CPU time for all four RTE orders at the initial test point. The first order (analytical) is excluded due to negligible CPU requirements. Subsequent RTE orders and MC calculations are presented with a precision of 1\% (relative uncertainty). One time unit in this context equates to 120 seconds of processing time on a system with 4 cores at 2223 MHz and 12 GB of RAM.
}
\label{tab:cpu_times}
\end{table}

We observe a remarkable agreement with the Monte-Carlo ray-tracing method for all four test points, providing strong validation for our approach. Extensive testing, including a large number of points, was conducted to ensure the reliability of our results. Our study reveals that RTE calculations, including single scattering and higher orders, are significantly faster than MC, with a speed advantage by tens or even thousands of times. The single scattering, being analytical, is virtually instantaneous and thus vastly more efficient than any MC computation. This efficiency persists even as we progress to higher-order RTE calculations, where despite the increased computational demand due to larger integral dimensions, RTE maintains a substantial lead in CPU time efficiency over MC, as detailed in~\cref{tab:cpu_times}.
In appendix~\labelcref{sec:test_orders}, we provide a detailed numerical analysis of the contributions from the first four scatterings. Key findings include the increased importance of multiple scatterings for side and backward test points compared to forward test points.

It's important to note that the Monte Carlo method faces limitations, particularly for distant points, where the probability of photons reaching a small detector sphere with a radius of $0.21$ meters from a distance of a hundred meter distance becomes exceedingly small.

\subsection{Analyzing the Impact of Scattering Counts on Signal Expectation}
\label{sec:quantify_orders}
\begin{figure}[!]
\centering
\includegraphics[width=\linewidth]{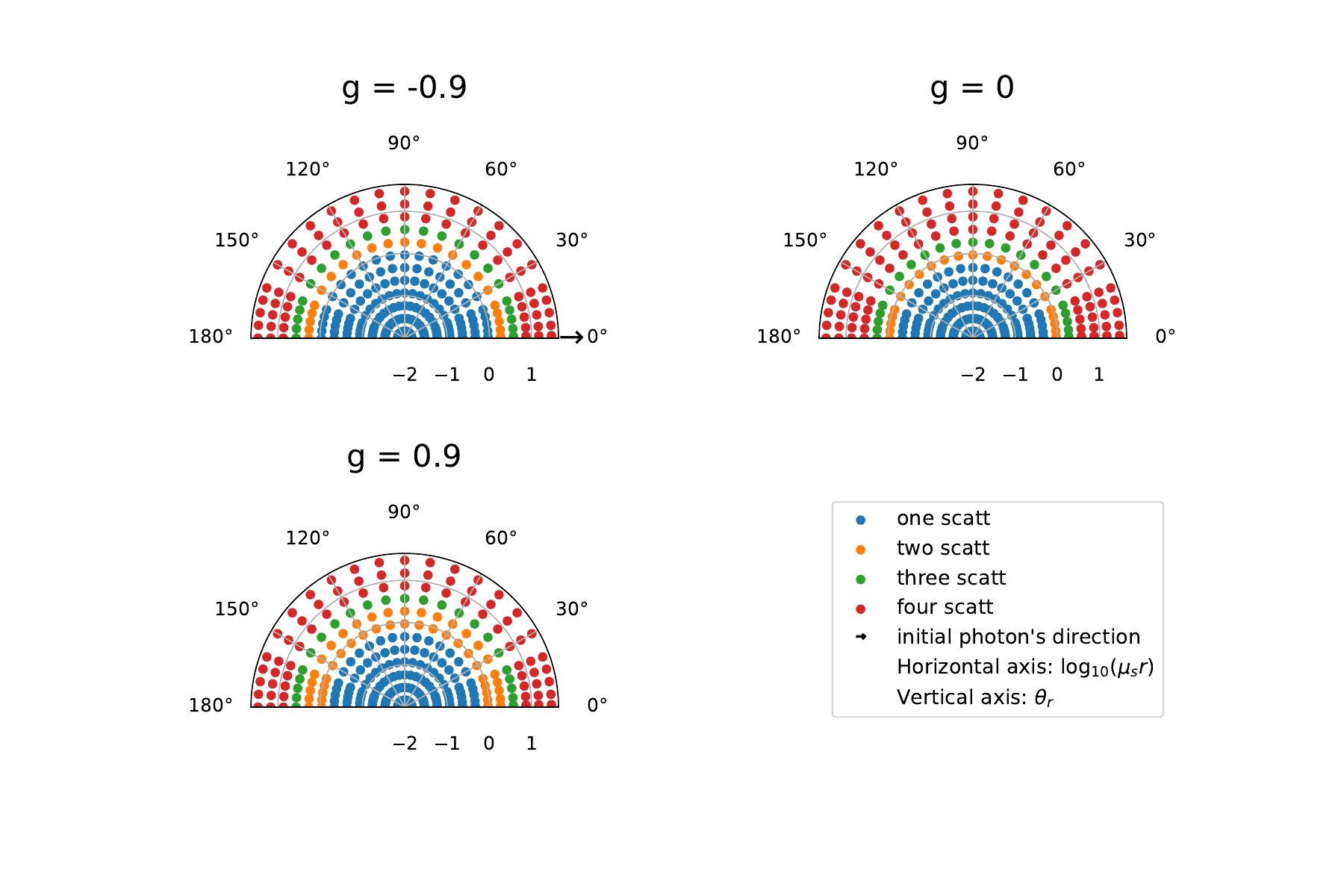}
\caption{Polar diagrams for $(r,\theta_r)$ illustrating the dominance of scattering order $n \in (1, 4)$. The diagrams are presented for three distinct values of the asymmetry parameter: $g=\pm 0.9$ and $g=0$, demonstrating the variation in scattering order dominance across different regimes of anisotropy.}
\label{fig:polar_plots}
\end{figure}
In this section, we continue our analysis utilizing the properties of the medium as outlined in~\cref{sec:comparison}. In particular, we delve into the crucial role that the number of scatterings plays in shaping the expected signal. To investigate this, we consider a spatial grid defined by the distance $r$ and zenith angle $\theta_r$, with the azimuthal angle $\varphi_r$ set to zero due to azimuthal symmetry. For each coordinate pair $(r, \theta_r)$, we compute the expected signal contributions from one to four scatterings. These individual contributions are then compared to identify the predominant scattering order. \Cref{fig:polar_plots} visually represents the domains in the $(r, \theta_r)$ space where specific scattering orders, ranging from one to four, are dominant.

Contrary to a naive expectation that multiple scatterings gain significance at $r \approx \mu_s^{-1}$, our findings reveal that the prevalence of different scattering orders is heavily influenced by distance, zenith angle, and the asymmetry parameter. Let us discuss these three cases in some more details.

Contrary to a naive expectation that multiple scatterings gain significance at $r \approx \mu_s^{-1}$, our findings reveal that the prevalence of different scattering orders is heavily influenced by distance and the asymmetry parameter. Let us discuss these three cases in some more details.

(i) In the case of $g \rightarrow 1$, the first scattering order predominates within a radius of $r < \mu_s^{-1}$. Beyond $r \gtrsim 4\mu_s^{-1}$, the third scattering order becomes preeminent. At lengths greater than 7.4$\mu_s^{-1}$, orders above the third order  predominate.

(ii) For $g \rightarrow -1$, the first scattering order is dominant within a radius of $r \lesssim 2 \mu_s^{-1}$, while the second and third orders dominate until  $r \lesssim 7.4 \mu_s^{-1}$.

(iii) In the isotropic scattering scenario ($g = 0$), the first scattering order is dominant within $r \gtrsim 0.85\mu_s^{-1}$ and the fourth order begins to dominate at $r \gtrsim 3.7\mu_s^{-1}$

\begin{figure}[!]
\centering
\includegraphics[width=\linewidth]{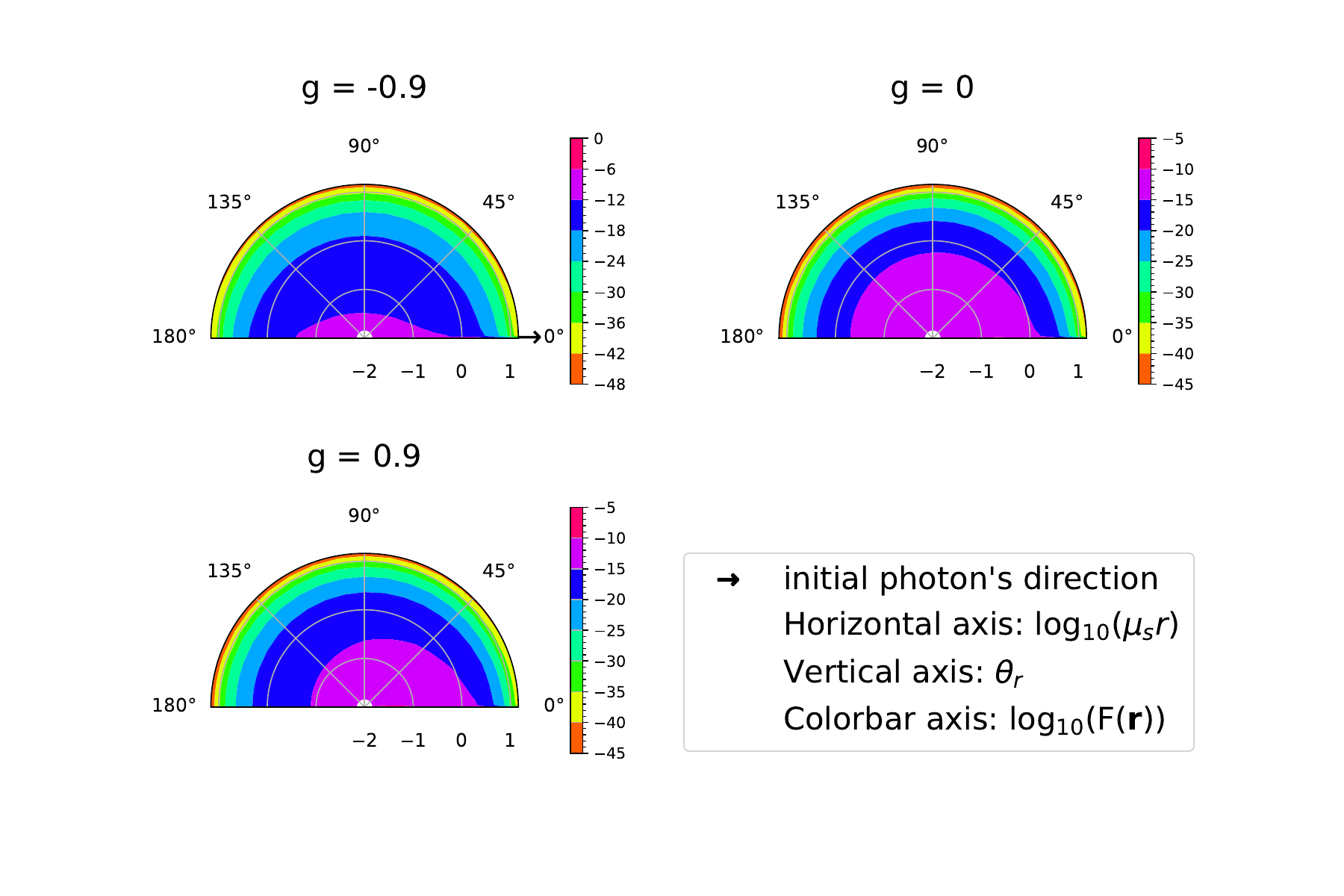}
\caption{Polar representation of fluence $F(\bm{r})$ as described in~\cref{eq:fluence}, considering absorption and scattering lengths $\mu_a^{-1} = 20.9$ m and $\mu_s^{-1} = 69.26$ m, accounting for four scattering orders. These diagrams display the fluence patterns for three distinct asymmetry parameter values: $g=\pm 0.9$ and $g=0$, elucidating the influence of different $g$ values on the spatial distribution of fluence.}
\label{fig:polar_plots_abs}
\end{figure}
In~\cref{fig:polar_plots_abs} we display fluence $F(\bm{r})$ as described in~\cref{eq:fluence}, assuming  absorption and scattering lengths $\mu_a^{-1} = 20.9$ m and $\mu_s^{-1} = 69.26$ m, accounting for four scattering orders. \Cref{fig:polar_plots,fig:polar_plots_abs} are instrumental in developing optimal strategies for practical calculations of photon flux. Such strategies must consider the computational intensity of multiple scatterings, which generally require more CPU time to accurately compute.

\begin{figure}[!]
\centering
\includegraphics[width=\linewidth]{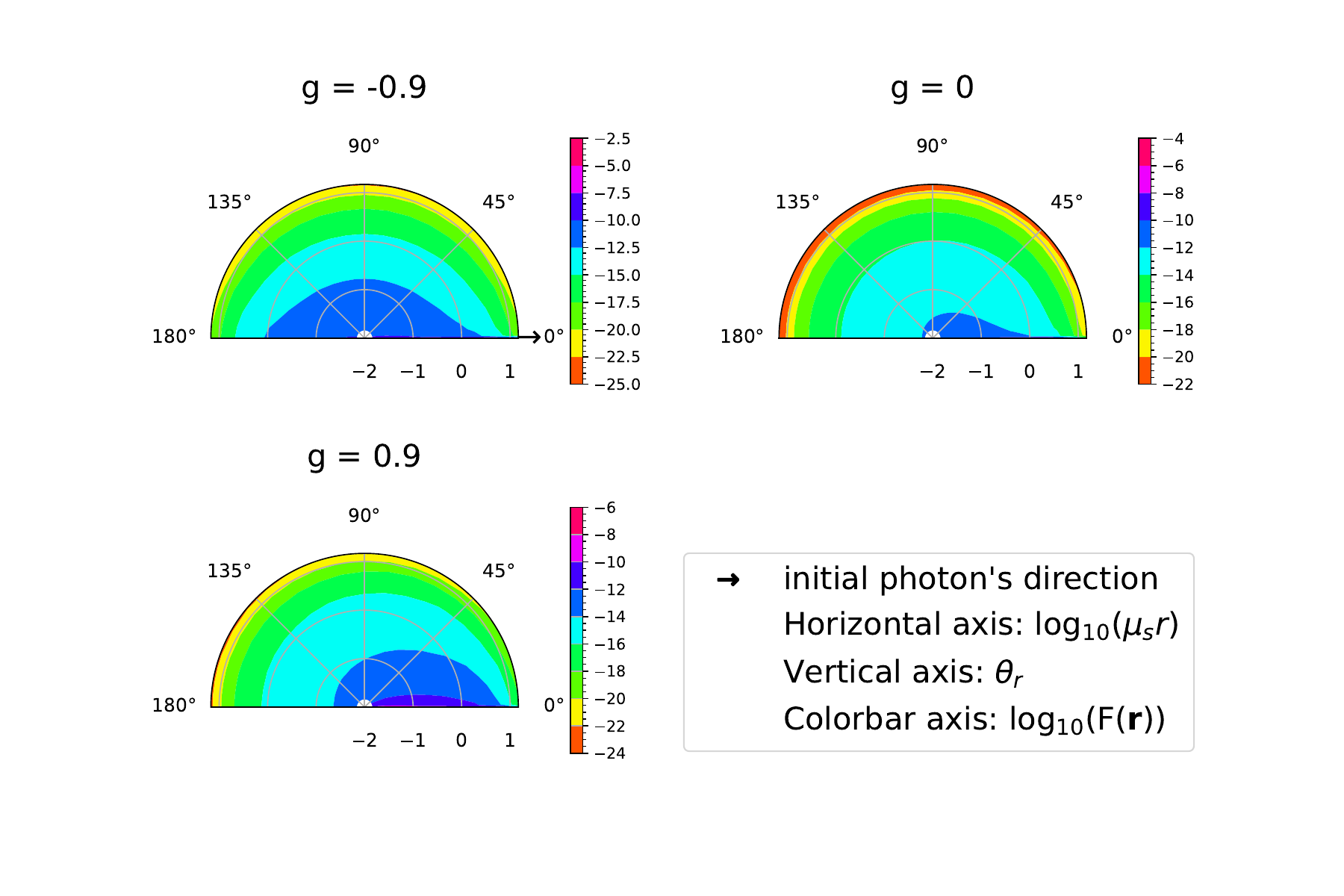}
\caption{Polar representation of fluence $F(\bm{r})$ as described in~\cref{eq:fluence}, assuming  $\mu_a = 0$, accounting for four scattering orders. These diagrams display the fluence patterns for three distinct asymmetry parameter values: $g=\pm 0.9$ and $g=0$, elucidating the influence of different $g$ values on the spatial distribution of fluence.}
\label{fig:polar_plots_noabs}
\end{figure}

As depicted in~\cref{fig:polar_plots_abs}, the fluence is influenced by the assumed absorption length. To emphasize the impact of scattering on fluence, ~\cref{fig:polar_plots_noabs} presents a scenario with an infinitely large absorption length. As one might intuitively anticipate, the presence of absorption contributes to the isotropization of light distribution, highlighting the interplay between absorption and scattering in shaping the fluence.
\section{Discussion and Conclusions}
\label{sec:conclusions}
Key findings of this work can be summarized as follows. We found a solution of the time-dependent linearized radiative transfer equation for a unidirectional instantaneous source as a series, where each term corresponds to a specific number of light scatterings. The zero-order term, known in literature as the ballistic solution and characterized by no scattering, is a singular function well-established in previous studies. The analytical solution for single scattering, found in this work, is also a singular function. Its singularity may pose challenges for certain numerical schemes aimed at resolving the time-dependent RTE, if not properly managed.

For higher scattering orders, our approach necessitates numerical integration of multi-dimensional integrals, incrementing three dimensions with each additional scattering event. The employment of the VEGAS integrator in our analysis effectively addresses this computational complexity, suggesting potential areas for future research to enhance the efficiency of multi-dimensional integral evaluations.

Our practical application analysis examines varying positions $\bm{r}$ and the asymmetry parameter $g$. This examination has delineated domains where single or double scatterings predominate, providing valuable insights for optimizing photon flux calculations.

When compared to Monte-Carlo ray-tracing, our solution represented by the first several orders show remarkable agreement. Importantly, our approach significantly enhances computational performance compared to the traditional Monte-Carlo method, featurning random end points, demonstrating its efficacy and potential for practical application.

Furthermore, the solutions derived here facilitate a nuanced analysis of light flux moments over time, clearly segregating the transport and diffusive light propagation regimes.
The exploitation of hidden symmetries, found in the course of this work, further accelerates our numerical calculations, enhancing the method's overall efficiency.

While the immediate application of our findings is within neutrino telescope analyses, the implications extend beyond, potentially benefiting various scientific and experimental domains where light propagation analysis is critical.

\section{Acknowledgements}
The authors extend special thanks to Yu.~Malyshkin and W.~Noga for their thorough cross-checks and validations of the results, and to V.~A.~Naumov for insightful discussions. We also express our gratitude to V.~A. Naumov, along with I. Belolaptikov, J.~M.~Djilkibaev, M.~Gonchar, L.~Kolupaeva, O.~Smirnov, and D.~Zaborov, for reviewing the manuscript and providing valuable suggestions that enhanced its quality.

\begin{appendices} 

\section{Transformation of RTE to the integral equation}
\label{sec:transform_to_integral}
A transformation of~\cref{eqref:RTE_0} to the integral equation can be done
%
by the following three steps.

(i) Transform the coordinates ($\bm{r},t$) to
\begin{equation}
    t \to t' =t, \bm{r} \to \bm{r}' = \bm{r} - c\bm{\hat{s}} t.
\end{equation}

(ii) Use the  identity
\begin{equation}
    \frac{\partial}{\partial t'} f(\bm{r}'+c\bm{\hat{s}}t',t') = c\bm{\hat{s}}\cdot \bm{\nabla} f(\bm{r},t) + \frac{\partial f(\bm{r},t)}{\partial t}.
\end{equation}

(iii) Re-write the left-hand-side of~\cref{eqref:RTE_0} as
\begin{equation}
 \label{eqref:RTE_3}
        \left(\frac{\partial}{c \partial t}+\hat{\bm{s}}\cdot\bm{\nabla}+\mu_t\right)L({\bm{r}},t,\bm{\hat{s}}) =e^{-c\mu_t t'}\frac{\partial}{\partial t'}\left[e^{c\mu_t t'}L(\bm{r}'+c\bm{\hat{s}}t',t',\bm{\hat{s}})\right].
\end{equation}

Integrating~\eqref{eqref:RTE_3} over $t'$ between $0$ and $t$ yields the integral equation
\begin{equation}
\begin{aligned}
    e^{c\mu_t t}L(\bm{r}'+c\bm{\hat{s}}t,t,\bm{\hat{s}}) &= L(\bm{r}',0,\bm{\hat{s}})
    +\int_{0}^{t} dt' c e^{c\mu_t t'}S({\bm{r}'}+c\bm{\hat{s}}t',t',\bm{\hat{s}})\\
    &+ c\mu_s\int_{0}^{t} dt' e^{c\mu_t t'}\hat{V}_{\bm{\hat{s}}\bm{\hat{s}}'}L({\bm{r}'+c\bm{\hat{s}}t'},t',\bm{\hat{s}}').
\end{aligned}
\end{equation}
One can put $L(\bm{r}',0,\bm{\hat{s}})=0$, assuming no light at $t=0$, or re-attributing it to the source function.

Returning back to the original variable $\bm{r} = \bm{r}' + c\bm{\hat{s}} t$ one gets~\cref{eqref:RTE_4} with definitions in~\cref{eq:RTE_integral}.

\section{Derivation of $\delta L^{(1)}$ in the Cartesian Coordinates}
\label{app:first_order_unidirectional_derivation}
Substituting~\cref{sec:source_delta} into the first line of~\cref{eq:RTE_integral} one gets:
\begin{equation}
    L_{0} = c e^{-\mu_t c t} \delta^3(\bm{r} - c t \hat{\bm{s}}_0) \delta^2(\hat{\bm{s}} - \hat{\bm{s}}_0).
\end{equation}
Substituting $L_0$ into the second line of~\cref{eq:RTE_integral} one gets:
\begin{equation}
\begin{aligned}
    \delta L^{(1)}
    &= \mu_s c \hat{V}_{\hat{\bm{s}} \hat{\bm{s}}'} \int\limits_{0}^{t} dt' e^{-c \mu_t (t-t')} L_0(\bm{r} - c \hat{\bm{s}}(t-t'), t', \hat{\bm{s}}')\\
    &=\mu_s c^2 \int\limits_{\mathbb{S}^2} d\hat{\bm{s}}_1 f(\hat{\bm{s}}\cdot\hat{\bm{s}}_1) \int\limits_{0}^{t} dt_1 e^{-\mu_t c t} \delta^3(\bm{r} - c (t - t_1) \hat{\bm{s}} - c t_1 \hat{\bm{s}}_0) \delta^2(\hat{\bm{s}}_1 - \hat{\bm{s}}_0)\\
    &= \mu_s c^2 f(\hat{\bm{s}}\cdot\hat{\bm{s}}_0) \int\limits_{0}^{t} dt_1 e^{-\mu_t c t} \delta^3(\bm{r} - c (t - t_1) \hat{\bm{s}} - c t_1 \hat{\bm{s}}_0)
\end{aligned}
\label{eq:eq:first_order_unidirectional_1}
\end{equation}
One has to find the root $t_1^*$ that zeros the argument of the Dirac delta function: $\delta\bm{r}=\bm{r} - c (t - t_1^*) \hat{\bm{s}} - c t_1^* \hat{\bm{s}}_0=0$. Assuming for simplicity $\hat{\bm{s}}_0 = (0, 0, 1)$, and expanding this equation along each axis, we have:
\begin{equation}
    \begin{aligned}
        x - c (t - t_1^*) \sin \theta \cos \varphi &= 0, \\
        y - c (t - t_1^*) \sin \theta \sin \varphi &= 0, \\
        z - c (t - t_1^*) \cos \theta - c t_1^*    &= 0.
    \end{aligned}
\label{eq:first_order_unidirectional_delta_argument}
\end{equation}
Introducing the polar distance $\rho = \sqrt{x^2 + y^2}$, one can get:
\begin{equation}
    \begin{aligned}
        \rho                &= c (t - t_1^*) \sin \theta, \\
        z - c t \cos \theta &= c t_1^* (1 - \cos \theta),
    \end{aligned}
\end{equation}
which allows one to find the root $t_1^*$:
\begin{equation}
    t_1^* = \frac{z - \rho \cot \theta}{c}, \text{ and } t - t_1^* = \frac{c t - z + \rho \cot \theta}{c}.
    \label{eq:t1_root}
\end{equation}
Therefore,
\begin{equation}
\begin{aligned}
    \delta^3(\delta\bm{r}) & = \frac{1}{\rho} \delta\left(z - c t \cos \theta - c t_1(1 - \cos \theta)\right) \times\\
                           & \times\delta(\rho - c (t - t_1) \sin \theta)\times \delta(\varphi - \varphi^*)
    \end{aligned}
\end{equation}
where $\varphi^*$ is determined by the first two lines of~\cref{eq:first_order_unidirectional_delta_argument}:
\begin{equation}
    \varphi^* = \arctan(y / x)
\end{equation}
Now the integration over $t_1$ in~\cref{eq:eq:first_order_unidirectional_1}  can be carried out using the root found  in~\cref{eq:t1_root}:
\begin{equation}
    \delta L^{(1)} = c\mu_s  e^{-\mu_t c t} \frac{f(\hat{\bm{s}}\cdot\hat{\bm{s}}_0)}{\rho (1 - \cos \theta)}  \delta(\rho - c (t - t_1^*) \sin \theta)\delta(\varphi - \varphi^*)H(ct-z)H(\frac{z}{ct}-\cos\theta),
\end{equation}
where the Heaviside functions $H$ are due to $t_1^*\ge 0$ and $t_1^*\le t$.

Using the second of equations in~\cref{eq:t1_root} one can simplify the argument of the first $\delta$-function:
\begin{equation}
        \rho - c (t - t_1^*) \sin \theta =  \rho - (c t - z) \cot \frac{\theta}{2},
\end{equation}
thus yielding~\cref{eq:first_order_unidirectional}.

\section{Derivation of $\delta L^{(1)}$ in the Vector Form}
\label{app:single_scatter_vector_form_derivation}
Let us apply the following equality
\begin{equation}
\delta^3(\bm{x}) = \frac{1}{|\bm{x}|^2}\delta(|\bm{x}|)\delta^2(\hat{\bm{x}}),
\end{equation}
where $\hat{\bm{x}}=\bm{x}/|\bm{x}|$, to $\delta^3\left(\delta\bm{r}\right)$ with $\delta\bm{r} = \bm{r} - c (t - t_1^*) \hat{\bm{s}} - c t_1^* \hat{\bm{s}}_0$:
\begin{equation}
    \delta^3\left(\delta\bm{r}\right) = \frac{1}{(c(t-t_1))^2} \delta\left(|\bm{r} - c\hat{\bm{s}}_0 t_1| - c(t-t_1)\right)\delta^2\left(\hat{\bm{s}} - \frac{\bm{r} - c\hat{\bm{s}}_0 t_1}{c(t-t_1)}\right)
    \label{eq:single_scatter_vector_form1}
\end{equation}
Let us find the time $t_1=t_1^*$ which zeros the argument of the first delta-function.
From here, we need to express the time $t_1$ from the second delta function. Let us denote
\begin{equation}
    g = |\bm{r} -c\hat{\bm{s}}_0 t_1| - c(t-t_1)= 0.
\end{equation}
Squaring this equation allows us to find its root and a couple of useful identities:
\begin{equation}
     t_1^* = \frac{c^2 t^2 - \bm{r}^2}{2c(\bm{r} \cdot \hat{\bm{s}}_0 - ct)}, \quad
     t - t_1^* = \frac{|\bm{r} - c\hat{\bm{s}}_0 t|^2}{2c(\bm{r} \cdot \hat{\bm{s}}_0 - ct)},
     \quad \frac{dg}{dt_1} = \frac{(c\hat{\bm{s}}_0 t-\bm{r}) \cdot \hat{\bm{s}}_0}{t-t_1}.
    \label{eq:single_scatter_vector_form2}
\end{equation}
Using first two of these equations allows us to express the directional unit vector $\bm{s}$ as follows:
\begin{equation}
    \begin{aligned}
        \hat{\bm{s}}^* &= \frac{\bm{r} - c\hat{\bm{s}}_0 t_1^*}{c(t-t_1^*)} = \frac{\bm{r} - c\hat{\bm{s}}_0 t + c\hat{\bm{s}}_0 (t-t_1^*)}{c(t-t_1^*)} = \hat{\bm{s}}_0 + \frac{\bm{r} - c\hat{\bm{s}}_0 t}{c(t-t_1^*)} \\
        &= \hat{\bm{s}}_0 + 2\frac{\bm{r} - c\hat{\bm{s}}_0 t}{|\bm{r} - c\hat{\bm{s}}_0 t|^2}(ct-\bm{r} \cdot \hat{\bm{s}}_0),
    \end{aligned}
    \label{eq:single_scatter_vector_form3}
\end{equation}
which is~\cref{eq:magic_direction}.

Next, combining~\cref{eq:single_scatter_vector_form1,eq:single_scatter_vector_form2,eq:single_scatter_vector_form3} yields:
\begin{equation}
\begin{aligned}
    \delta^3(\delta\bm{r}) &= \frac{1}{(c(t-t_1))^2} \delta\left(|\bm{r} - c\hat{\bm{s}}_0 t_1| - c(t-t_1)\right)\delta^2\left(\hat{\bm{s}} - \frac{\bm{r} - c\hat{\bm{s}}_0 t_1}{c(t-t_1)}\right) \\
    &= \frac{1}{c^2(t-t_1)(ct - \bm{r} \cdot \hat{\bm{s}}_0)} \delta^2\left(\hat{\bm{s}} - \hat{\bm{s}}^*\right) \delta(t_1-t_1^*)\\
    &= \frac{2}{c|\bm{r} - c\hat{\bm{s}}_0 t|^2} \delta^2\left(\hat{\bm{s}} - \hat{\bm{s}}^*\right) \delta(t_1-t_1^*)
\end{aligned}
\label{eq:single_scatter_vector_form_delta2}
\end{equation}
Now the integration over $t_1$ is straightforward yielding~\cref{eq:single_scatter_vector_form}, where the product of two Heaviside functions appears due to $t_1^* \ge 0$ and $t_1^* \le t$.

Finally, let us derive~\cref{eq:time_single_scattering}. The argument $\hat{\bm{s}} - \frac{\bm{r} - c\hat{\bm{s}}_0 t_1}{c(t-t_1)}$ of the angular delta function in~\cref{eq:single_scatter_vector_form_delta2}
 can be recast in the form
\begin{equation}
    \bm{r} - c\hat{\bm{s}}_0 t = c t_1 (\hat{\bm{s}}_0 - \hat{\bm{s}}).
    \label{eq:single_scatter_vector_form_delta2_recast}
\end{equation}
Multiplying this equation by $\hat{\bm{s}}_0 + \hat{\bm{s}}$, one can get:
\begin{equation}
    (\bm{r} - c\hat{\bm{s}}_0 t) \cdot (\hat{\bm{s}}_0 + \hat{\bm{s}}) = 0,
\end{equation}
yielding the second equation of~\cref{eq:time_single_scattering}.

It remains to prove the first equation of~\cref{eq:time_single_scattering}. While $t_1^*$ was already found in~\cref{eq:single_scatter_vector_form2} as a function of $t$ and $\hat{\bm{s}}_0$ it is possible reexpress it as function of $\hat{\bm{s}}_0$
and $\hat{\bm{s}}$ because $t$, $\hat{\bm{s}}_0$ and $\hat{\bm{s}}$ are related to each other for the single scattering.
To this end we multiply~\cref{eq:single_scatter_vector_form_delta2_recast} by $\hat{\bm{s}}_0$ to obtain:
\begin{equation}
    t = \frac{(\bm{r} \cdot (\hat{\bm{s}}_0 + \hat{\bm{s}}))}{c(1+(\hat{\bm{s}} \cdot \hat{\bm{s}}_0))}
\end{equation}
Then,~\cref{eq:single_scatter_vector_form_delta2_recast} can be re-written in two equivalent forms:
\begin{equation}
\begin{aligned}
    c(t-t_1)\hat{\bm{s}} &= \bm{r} - c\hat{\bm{s}}_0 t_1,\\
    c(t-2t_1)\hat{\bm{s}} &= \bm{r} - c(\hat{\bm{s}}_0+\hat{\bm{s}})t_1,
\end{aligned}
\end{equation}
which after multiplication by $\hat{\bm{s}}_0 - \hat{\bm{s}}$, gives us:
\begin{equation}
    c(t-2t_1)(\hat{\bm{s}} \cdot \hat{\bm{s}}_0 - \hat{\bm{s}}) = (\bm{r} \cdot \hat{\bm{s}}_0 - \hat{\bm{s}}).
\end{equation}
From here, we get $t_1$ as function of $\hat{\bm{s}}_0$
and $\hat{\bm{s}}$
\begin{equation}
    t_1 = \frac{1}{2c} \left(\frac{\bm{r} \cdot (\hat{\bm{s}}_0 - \hat{\bm{s}})}{1 - \hat{\bm{s}} \cdot \hat{\bm{s}}_0}
    + \frac{\bm{r} \cdot (\hat{\bm{s}}_0 + \hat{\bm{s}})}{1 + \hat{\bm{s}} \cdot \hat{\bm{s}}_0}\right)
\end{equation}

\section{Treatment of time integrals}
\label{sec:time_integrals}
The time integrals in~\cref{eq:deltan_0} could be significantly simplified if time variables $t_k$ are replaced by  dimensionless $\xi_k = 1 - \frac{t_k}{t_{k+1}}$. Then,
\begin{equation}
    dt_1\ldots dt_n =  |J_n| d\xi_1\ldots d\xi_n,
\end{equation}
where $J_n$ is the determinant
\begin{equation}
    J_n = \frac{\partial(t_1,\ldots,t_n)}{\partial(\xi_1,\ldots,\xi_n)}.
\end{equation}
It is simpler to calculate first its inverse, $J_n^{-1}=\det{\frac{\partial \xi_j}{\partial t_k}}$ with
\begin{equation}
    \frac{\partial\xi_k}{\partial t_k} = -\frac{1}{t_{k+1}}, \frac{\partial\xi_k}{\partial t_{k+1}}=\frac{t_k}{t^2_{k+1}}.
\end{equation}
Explicit form
\begin{equation}
    J^{-1}_n =
\begin{vmatrix}
  -\frac{1}{t_2} & \frac{t_1}{t^2_2} & 0 & 0& \ldots & 0\\
  0 & -\frac{1}{t_3} & \frac{t_2}{t^2_3} & 0 &\ldots & 0\\
  0 & 0& -\frac{1}{t_4} & \frac{t_3}{t^2_4}  &\ldots & 0\\
  \ldots & \ldots & \ldots & \ldots &\ldots & \ldots\\
  0 & 0 & 0 & \ldots & -\frac{1}{t_n}& \frac{t_n}{t^2_{n+1}}\\
  0 & 0 & 0 &  \ldots &0& -\frac{1}{t_{n+1}}\\
\end{vmatrix}
\end{equation}
allows us to find
\begin{equation}
\label{eq:inverse_J}
    J^{-1}_n = \frac{(-1)^n}{t_{n+1}t_n\ldots t_3t_2} = \frac{(-1)^n}{t t_n\ldots t_3t_2}
\end{equation}
$J^{-1}_n$ has to be re-written in terms of $\xi_j$ variables.
In order to express the time variables $t_k$ via $\xi_j$ we notice
\begin{equation}
    \begin{aligned}
        t_n     & = t_{n+1}(1-\xi_n) = t(1-\xi_n)\\
        t_{n-1} & = t_n(1-\xi_{n-1}) = t(1-\xi_n)(1-\xi_{n-1}),
    \end{aligned}
\end{equation}
which allows us to find a general formula
\begin{equation}
\label{eq:time_vs_xi}
    \begin{aligned}
        t_k     & = t_{k+1}(1-\xi_k) = t_{k+2}(1-\xi_k)(1-\xi_{k+1})\\
                & = t(1-\xi_n)(1-\xi_{n-1})\ldots(1-\xi_k).
    \end{aligned}
\end{equation}
Inserting $t_k$ from~\cref{eq:time_vs_xi} into $J^{-1}_n$ in~\cref{eq:inverse_J} and inverting it, yields
\begin{equation}
    J_n = t^n\prod\limits_{k=1}^{n}(1-\xi_k)^{k-1}.
\end{equation}
The following identity
\begin{equation}
\begin{aligned}
    t_{k+1}-t_k & = t_{k+1}\xi_k\\
                & = t(1-\xi_n)(1-\xi_{n-1})\ldots(1-\xi_{k+1})\xi_k
\end{aligned}
\end{equation}
 is useful to re-write the argument of the delta-function in~\cref{eq:deltan_0} as $\bm{r} - ct\bm{s}^{(n)}$ with $\bm{s}^{(n)}$ given by~\cref{eq:s_n}.

\section{Convolution Property of the Scattering Function and Its Asymmetry Parameter}
\label{sec:Henyey-Greenstein-properties}
The Henyey-Greenstein scattering function, denoted as $f(\hat{\bm{s}} \cdot \hat{\bm{s}}')$, is dependent on the asymmetry parameter $g$.
Throughout this section, we will emphasize this dependence by introducing a subscript, thus representing it as $f_g(\hat{\bm{s}} \cdot \hat{\bm{s}}')$:
\begin{equation}
f_g(\hat{\bm{s}} \cdot \hat{\bm{s}}') = \frac{1}{4\pi}\frac{1-g^2}{\left(1+g^2-2g(\hat{\bm{s}} \cdot \hat{\bm{s}}')\right)^{3/2}}.
\label{eq:Henyey-Greenstein}
\end{equation}
In the foundational study by Henyey and Greenstein~\cite{henyey1941diffuse}, the function $f_g(\hat{\bm{s}} \cdot \hat{\bm{s}}')$ is also influenced by the spherical albedo factor $\gamma$. In our analysis, we simplify this by setting $\gamma=1$.

A remarkable property of  function in~\cref{eq:Henyey-Greenstein} is its convolution property, which can be expressed as:
\begin{equation}
\int\limits_{\bm{S}^2}d\hat{\bm{s}}' f_g(\hat{\bm{s}}_1\cdot \hat{\bm{s}}')f_g(\hat{\bm{s}}'\cdot \hat{\bm{s}}_2) = f_{g^2}(\hat{\bm{s}}_1\cdot \hat{\bm{s}}_2).
\label{eq:hg_convolution_1}
\end{equation}
The convolution of the Henyey-Greenstein scattering function with an asymmetry parameter $g$ yields another Henyey-Greenstein scattering function with the asymmetry parameter squared, i.e., $g^2$. A possible interpretation suggests that the effect of scatterings is to smear the scattering function.

The property described in~\cref{eq:hg_convolution_1} can be generalized to consider the convolution of two functions with distinct asymmetry parameters:
\begin{equation}
\int\limits_{\bm{S}^2}d\hat{\bm{s}}'f_{g_1}(\hat{\bm{s}}_1\cdot \hat{\bm{s}}')f_{g_2}(\hat{\bm{s}}'\cdot \hat{\bm{s}}_2) = f_{g_1g_2}(\hat{\bm{s}}_1\cdot \hat{\bm{s}}_2).
\label{eq:hg_convolution_2}
\end{equation}
Since in the limit $g\to 1$, the Henyey-Greenstein scattering function approaches the Dirac delta-function:
\begin{equation}
\lim\limits_{g\to 1}f_{g}(\hat{\bm{s}}_1\cdot \hat{\bm{s}}_2) = \delta^2(\hat{\bm{s}}_1-\hat{\bm{s}}_2),
\label{eq:hg_convolution_3}
\end{equation}
\cref{eq:hg_convolution_2,eq:hg_convolution_3} allow us to derive an intriguing relationship:
\begin{equation}
\int\limits_{\bm{S}^2}d\hat{\bm{s}}'f_{g}(\hat{\bm{s}}_1\cdot \hat{\bm{s}}')f_{1/g}(\hat{\bm{s}}'\cdot \hat{\bm{s}}_2) = \delta^2(\hat{\bm{s}}_1-\hat{\bm{s}}_2).
\label{eq:hg_convolution_4}
\end{equation}
Please, note that $f_{1/g}(x)$ should be regarded as a generalized function and does not have a direct interpretation as a probability density function.

To establish the fundamental relationship in~\cref{eq:hg_convolution_2}, we can employ the well-known decomposition:
\begin{equation}
f_{g}(\hat{\bm{s}}\cdot \hat{\bm{s}}') = \sum\limits_{l=0}^\infty \sum\limits_{m=-l}^l g^l Y_{lm}(\hat{\bm{s}})Y^*_{lm}(\hat{\bm{s}}').
\label{eq:hg_convolution_4a}
\end{equation}
Therefore,
\begin{equation}
\begin{aligned}
\int\limits_{\bm{S}^2}d\hat{\bm{s}}'&f_{g_1}(\hat{\bm{s}}_1\cdot \hat{\bm{s}}')f_{g_2}(\hat{\bm{s}}'\cdot \hat{\bm{s}}_2)
=\\
&=\sum\limits_{l=0}^\infty \sum\limits_{m=-l}^l
\sum\limits_{l'=0}^\infty \sum\limits_{m'=-l'}^{l'}
g_1^{l}g_2^{l'}Y_{lm}(\hat{\bm{s}}_1)Y^*_{l'm'}(\hat{\bm{s}}_2)\int\limits_{\bm{S}^2}d\hat{\bm{s}}'Y_{lm}(\hat{\bm{s}}')Y^*_{l'm'}(\hat{\bm{s}}')\\
&=\sum\limits_{l=0}^\infty \sum\limits_{m=-l}^l(g_1g_2)^l Y_{lm}(\hat{\bm{s}}_1)Y^*_{lm}(\hat{\bm{s}}_2)
= f_{g_1g_2}(\hat{\bm{s}}_1\cdot \hat{\bm{s}}_2).
\end{aligned}
\label{eq:hg_convolution_5}
\end{equation}
In the second line of~\cref{eq:hg_convolution_5} the orthogonality condition for spherical harmonics was employed, which states: $$\int\limits_{\bm{S}^2}d\hat{\bm{s}}'Y_{lm}(\hat{\bm{s}}')Y^*_{l'm'}(\hat{\bm{s}}')={\delta_{ll'}\delta_{mm'}}.$$

\section{First and Second Moments of the Scattering Function}
\label{app:moments_HG}
Let us derive the moments of the Henyey-Greenstein distribution given by~\cref{eq:Henyey-Greenstein}.
These moments are defined as follows:
\begin{equation}
    m_{ijk\ldots} = \int\limits_{\mathbb{S}^2} d\hat{\bm{s}}_1 f_g(\hat{\bm{s}}, \hat{\bm{s}}_1) \hat{\bm{s}}_{1i} \hat{\bm{s}}_{1j} \hat{\bm{s}}_{1k} \ldots \equiv \langle \hat{\bm{s}}_{1i} \hat{\bm{s}}_{1j} \hat{\bm{s}}_{1k} \ldots \rangle_{HG},
    \label{eq:moments_definition}
\end{equation}
where indices correspond to spatical coordinates.
Zeroth moment is just unity due to the normalization condition of the scattering function:
\begin{equation}
    \int\limits_{\mathbb{S}^2} d\hat{\bm{s}}_1 f_g(\hat{\bm{s}}, \hat{\bm{s}}_1) = 1.
\end{equation}
First-order moments are calculated as follows.
\begin{equation}
    m_{i} = \int\limits_{\mathbb{S}^2} d\hat{\bm{s}}_1 f_g(\hat{\bm{s}}, \hat{\bm{s}}_1) \hat{\bm{s}}_{1i} =a \hat{\bm{s}}_i,
    \label{eq:first_moment_HG_1}
\end{equation}
where the second equality uses the observation that the integral depends upon the single vector $\hat{\bm{s}}$. To find an unknown normalization constant $a$ we multilpy~\cref{eq:first_moment_HG_1}  by $\hat{\bm{s}}_i$ and sum over $i$:
\begin{equation}
    a = \int\limits_{\mathbb{S}^2} d\hat{\bm{s}}_1 f_g(\hat{\bm{s}}, \hat{\bm{s}}_1) (\hat{\bm{s}}, \hat{\bm{s}}_1) = 2\pi \int\limits_{-1}^{1} dx x f_g(x) = g.
\end{equation}
Thus, we get:
\begin{equation}
    m_{i} = \langle \hat{\bm{s}}_{1i}\rangle_{HG} = \int\limits_{\mathbb{S}^2} d\hat{\bm{s}}_1 f(\hat{\bm{s}}, \hat{\bm{s}}_1) \hat{\bm{s}}_{1i}  = g \hat{\bm{s}}_{i}.
\end{equation}
Finally, let us calculate the second moments, following its definition and seeking for its solution in the form:
\begin{equation}
    m_{ij} = \int\limits_{\mathbb{S}^2} d\hat{\bm{s}}_1 f(\hat{\bm{s}}, \hat{\bm{s}}_1) \hat{\bm{s}}_{1i} \hat{\bm{s}}_{1j}
    =a \delta_{ij} + b \hat{\bm{s}}_{i} \hat{\bm{s}}_{j}.
    \label{eq:second_moment_HG_1}
\end{equation}
Setting $i = j$ and summing over $i$, yields the first equation:
\begin{equation}
    3a + b = 1.
    \label{eq:second_moment_HG_2}
\end{equation}
Multiplying by $\hat{\bm{s}}_i \hat{\bm{s}}_j$ and summing over $i,j$, yields the second equation:
\begin{equation}
    a + b =  \int\limits_{\mathbb{S}^2} d\hat{\bm{s}}_1 f_g(\hat{\bm{s}}, \hat{\bm{s}}_1) (\hat{\bm{s}}, \hat{\bm{s}}_1)^2 = 2\pi \int\limits_{-1}^{1} dx x^2 f_g(x).
    \label{eq:second_moment_HG_3}
\end{equation}
Determining $a$ and $b$ from~\cref{eq:second_moment_HG_2,eq:second_moment_HG_3} and substituting them into~\cref{eq:second_moment_HG_1}, one can get:
\begin{equation}
    m_{ij} = \frac{1-g^2}{3} \delta_{ij} + g^2 \hat{\bm{s}}_{i} \hat{\bm{s}}_{j}.
    \label{disp:HG}
\end{equation}

\section{Chained First and Second Moments of the Scattering Function}
\label{app:moments_n_scattering_HG}
Here we consider the following integrals, similar to~\cref{eq:moments_definition}, but for a chain of the scattering functions:
\begin{equation}
\begin{aligned}
    M^{k}_i &= \int\limits_{\mathbb{S}^2} d\hat{\bm{s}}_1 \ldots \int\limits_{\mathbb{S}^2} d\hat{\bm{s}}_n f_g(\hat{\bm{s}}_0 \cdot \hat{\bm{s}}_1) \ldots f_g(\hat{\bm{s}}_{n-1} \cdot \hat{\bm{s}}_n) \left(\hat{\bm{s}}_k\right)_i,\\
    M^{km}_{ij} &= \int\limits_{\mathbb{S}^2} d\hat{\bm{s}}_1 \ldots \int\limits_{\mathbb{S}^2} d\hat{\bm{s}}_n f_g(\hat{\bm{s}}_0 \cdot \hat{\bm{s}}_1) \ldots f_g(\hat{\bm{s}}_{n-1} \cdot \hat{\bm{s}}_n) (\hat{\bm{s}}_k)_i (\hat{\bm{s}}_m)_j.
\end{aligned}
\label{eq:moments_definition_chained}
\end{equation}
To clarify the notation, $\left(\hat{\bm{s}}_k\right)_i$ denotes the $i$-th spatial component of the integration variable unit vector $\hat{\bm{s}}_k$. There are $n$ integrals over the unit vectors $\hat{\bm{s}}_1, \dots, \hat{\bm{s}}_n$. Therefore, $i \in [1, 3]$ and $k \in [1, n]$. Hence, the left-hand side of~\cref{eq:moments_definition_chained} depends on the integration variable number $k$, its spatial component $i$, and the total number of unit vectors $n$.

Please, note, that we explicitly specify the asymmetry parameter $g$ of the scattering function as its subscript.
Considering the convolution property~\cref{eq:hg_convolution_1} and normalization $\int\limits_{\mathbb{S}^2} d\hat{\bm{s}}_1 f_{g}(\hat{\bm{s}} \cdot \hat{\bm{s}}_1)=1$, one can obtain:
\begin{equation}
\begin{aligned}
    M^{k}_i &= g^k (\hat{\bm{s}}_0)_i, \text{ and 3d vector }\bm{M}^{k} = g^k\hat{\bm{s}}_0,\\
    M^{km}_{ij}&= g^{m-k} \left(\frac{1-g^{2k}}{3} \delta_{ij} + g^{2k} (\hat{\bm{s}}_0)_i (\hat{\bm{s}}_0)_j\right).
\end{aligned}
\label{eq:moments_definition_chained_results}
\end{equation}
To derive the second line of~\cref{eq:moments_definition_chained_results} the following tips might be helpful.
\begin{equation}
\begin{aligned}
    M^{km}_{ij} & = \int\limits_{\mathbb{S}^2} d\hat{\bm{s}}_1 \ldots \int\limits_{\mathbb{S}^2} d\hat{\bm{s}}_n f_g(\hat{\bm{s}}_0 \cdot \hat{\bm{s}}_1) \ldots f_g(\hat{\bm{s}}_{n-1} \cdot \hat{\bm{s}}_n) (\hat{\bm{s}}_k)_i (\hat{\bm{s}}_m)_j\\
    &= \int\limits_{\mathbb{S}^2} d\hat{\bm{s}}_k \ldots \int\limits_{\mathbb{S}^2} d\hat{\bm{s}}_m f_{g^k}(\hat{\bm{s}}_0 \cdot \hat{\bm{s}}_k) \ldots f_g(\hat{\bm{s}}_{m-1} \cdot \hat{\bm{s}}_m) (\hat{\bm{s}}_k)_i (\hat{\bm{s}}_m)_j\\
    &= g^{m-k} \left(\frac{1-g^{2k}}{3} \delta_{ij} + g^{2k} (\hat{\bm{s}}_0)_i (\hat{\bm{s}}_0)_j\right),
\end{aligned}
\end{equation}
where $k < m$ was assumed, the integration $m-k-1$ times was performed in the second line and~\cref{disp:HG} was used in the third line.

\section{Nested time integrals}
\label{app:nested_time_integrals}
Here we calculate two types of integrals, which are crucial in calculations of the moments of the light flux.

\noindent (i) The first type is defined as:
\begin{equation}
\begin{aligned}
    J_k &= \int\limits_0^{t}dt_n...\int\limits_0^{t_2}dt_1 (t_{k+1}-t_k)\\
        &= t^{n+1}\int\limits_0^{1}d\xi_n (1-\xi_n)^{n-1} \dots\int\limits_0^{1}d\xi_2(1-\xi_2) \int\limits_0^{1}d\xi_1  (t_{k+1}-t_k)\\
        &=t^{n+1}\int\limits_0^{1}d\xi_n (1-\xi_n)^{n} \dots\int\limits_0^{1}d\xi_{k+1} (1-\xi_{k+1})^{k+1}\times\\
        &\times\int\limits_0^{1}d\xi_k \xi_k(1-\xi_{k})^{k-1}\int\limits_0^{1}d\xi_{k-1} (1-\xi_{k-1})^{k-2}...\int\limits_0^{1}d\xi_2(1-\xi_2) \int\limits_0^{1}d\xi_1 \\
        &=t^{n+1}B(1,n+1)...B(1,k+2)B(2,k)B(1,k-1)\dots B(1,2)B(1,1) \\
        &=\frac{t^{n+1}B(2,k)}{B(1,k)B(1,k+1)}\prod\limits_{i = 1}^{n+1}B(1,i) = t^{n+1}\prod\limits_{i = 1}^{n+1}\frac{1}{i}  = \frac{t^{n+1}}{(n+1)!},
\end{aligned}
\end{equation}
where change of variables $\xi_i = 1-\frac{t_i}{t_{i+1}}$ is done in the second line and $B(p,q)$ is the Euler $\beta$-function for which
\begin{equation}
\begin{aligned}
    B(1,i) &= \frac{\Gamma(1)\Gamma(i)}{\Gamma(i+1)} = \frac{(i-1)!}{i!} = \frac{1}{i},\\
    B(2,k) &= \frac{\Gamma(2)\Gamma(k)}{\Gamma(k+2)} = \frac{(k-1)!}{(k+1)!} = \frac{1}{k(k+1)}.
\end{aligned}
\end{equation}
\noindent (ii) The second type is defined as follows, assuming $k<m$,
\begin{equation}
\begin{aligned}
    J_{km} &= \int\limits_0^{t}dt_n...\int\limits_0^{t_2}dt_1 (t_{k+1}-t_k)(t_{m+1}-t_m)\\
    &= t^{n+2}\int\limits_0^{1}d\xi_n (1-\xi_n)^{n+1} \dots\int\limits_0^{1}d\xi_{k+1} (1-\xi_{k+1})^{k+2}\times\\
    &\times\int\limits_0^{1}d\xi_k \xi_k(1-\xi_{k})^{k}\int\limits_0^{1}d\xi_{k-1} (1-\xi_{k-1})^{k-1}\dots
    \int\limits_0^{1}d\xi_{m+1} (1-\xi_{m+1})^{m+1}\times\\
    &\times\int\limits_0^{1}d\xi_m  \xi_m(1-\xi_{m})^{m-1}\int\limits_0^{1}d\xi_{m-1} (1-\xi_{m-1})^{m-2}\dots\int\limits_0^{1}d\xi_2(1-\xi_2) \int\limits_0^{1}d\xi_1  \\
    &= t^{n+2}B(1,n+2)\dots B(1,k+3)B(2,k+1)B(1,k)\dots B(1,m+2) \\
    &= B(2,m)B(1,m-1)\dots B(1,2)B(1,1) =  \frac{t^{n+2}B(2,k+1)}{B(1,k+1)B(1,k+2)}\times \\
    &\times\frac{B(2,m)}{B(1,m)B(1,m+1)}\prod\limits_{i = 1}^{n+2}B(1,i) = t^{n+2}\prod\limits_{i = 1}^{n+2}\frac{1}{i}  = \frac{t^{n+2}}{(n+2)!} \equiv J_{00},
\end{aligned}
\end{equation}
where the same change of variables $\xi_i = 1-\frac{t_i}{t_{i+1}}$ is done.

Finally, let us consider the case $k = m$:
\begin{equation}
\begin{aligned}
    J_{kk} &= t^{n+2}\int\limits_0^{1}d\xi_n (1-\xi_n)^{n+1}  ...\int\limits_0^{1}d\xi_{k+1} (1-\xi_{k+1})^{k+2}\times\\
    &\times\int\limits_0^{1}d\xi_k \xi_k^2(1-\xi_{k})^{k-1}\int\limits_0^{1}d\xi_{k-1} (1-\xi_{k-1})^{k-2}\dots
    \int\limits_0^{1}d\xi_2(1-\xi_2) \int\limits_0^{1}d\xi_1  \\
    &= t^{n+2}B(1,n+2)...B(1,k+3)B(3,k)B(1,k-1)...B(1,2)B(1,1) \\
    &= \frac{t^{n+2}B(3,k)}{B(1,k)B(1,k+1)B(1,k+2)}\prod\limits_{i = 1}^{n+2}B(1,i) = 2t^{n+2}\prod\limits_{i = 1}^{n+2}\frac{1}{i}  = 2\frac{t^{n+2}}{(n+2)!},
\end{aligned}
\end{equation}
where in the last line we accounted that
\begin{equation}
    B(3,k) = \frac{\Gamma(3)\Gamma(k)}{\Gamma(k+3)} = \frac{2(k-1)!}{(k+2)!} = \frac{2}{k(k+1)(k+2)}.
\end{equation}
Therefore, one can get
\begin{equation}
    J_{km} = J_{00}(1+\delta_{km}).
    \label{eq:double_momentum_for_time}
\end{equation}

\section{Derivation of the Moments of Light Flux}
\label{app:moments_light_flux}
Here we use the definition of the average as given by~\cref{eq:first_moment_definition,eq:first_moment_normalization_definition}. It is handy to define also a contribution to the average of an observable $O(\bm{r}, t, \hat{\bm{s}})$ due to no scattering and exactly $n \in [1,\infty]$ scattering events:
\begin{equation}
\begin{aligned}
    \langle \delta O(t) \rangle^{(0)} &= \frac{1}{N}\int\limits_{\mathbb{S}^2} d\hat{\bm{s}} \int d^3\bm{r} O(\bm{r}, t, \hat{\bm{s}}) L^{(0)}(\bm{r}, t, \hat{\bm{s}}),\\
    \langle \delta O(t) \rangle^{(n)} &= \frac{1}{N}\int\limits_{\mathbb{S}^2} d\hat{\bm{s}} \int d^3\bm{r} O(\bm{r}, t, \hat{\bm{s}}) \delta L^{(n)}(\bm{r}, t, \hat{\bm{s}}),
\end{aligned}
\end{equation}
where $N$ is given by~\cref{eq:first_moment_normalization_definition}, $L^{(0)}$ is given by~\cref{eq:zero_order_unidirectional} and $\delta L^{(n)}$ is given by~\cref{eq:deltan_0}.

To calculate the normalization $N(t)$, it is handy to observe that
\begin{equation}
\begin{aligned}
    \int\limits_{\mathbb{S}^2} d\hat{\bm{s}} \int d^3\bm{r} L^{(0)}(\bm{r}, t, \hat{\bm{s}}) &= c e^{-c\mu_t t},\\
    \int\limits_{\mathbb{S}^2} d\hat{\bm{s}} \int d^3\bm{r} \delta L^{(n)}(\bm{r}, t, \hat{\bm{s}}) &= c (\mu_s c t)^n e^{-c\mu_t t},
\end{aligned}
\end{equation}
which allows us to prove~\cref{eq:first_moment_normalization_definition}:
\begin{equation}
    N(t) =  \int\limits_{\mathbb{S}^2} d\hat{\bm{s}} \int d^3\bm{r} \left[L^{(0)}(\bm{r}, t, \hat{\bm{s}})+\sum\limits_{n=1}^{\infty}\delta L^{(n)}(\bm{r}, t, \hat{\bm{s}})\right] = c e^{-c\mu_a t}.
\end{equation}
Therefore,
\begin{equation}
\begin{aligned}
    \langle \delta O(t) \rangle^{(0)} &= e^{-c\mu_a t}O(ct\hat{\bm{s}}_0,t,\hat{\bm{s}}_0),\\
    \langle \delta O(t) \rangle^{(n)} &= (\mu_s c)^n e^{-\mu_s c t} \int d\bm{r} \left(\prod_{i=1}^n \int\limits_0^{t_{i+1}} dt_i\right) \left(\prod_{j=1}^{n} \int\limits_{\mathbb{S}^2} d\hat{\bm{s}}_j f(\hat{\bm{s}}_{j} \cdot \hat{\bm{s}}_{j-1})\right) \times\\
    &\times
    O(\bm{r}, t, \hat{\bm{s}}) \delta^3(\bm{r} - c \sum_{i=0}^n \hat{\bm{s}}_i (t_{i+1} - t_i))
\end{aligned}
\end{equation}
and the average reads:
\begin{equation}
    \langle O(t) \rangle = \sum\limits_{n=0}^{\infty} \langle \delta O(t) \rangle^{(n)}
\end{equation}

\subsection{Mean coordinate}
\label{app:moments_light_flux_mean_coordinate}
We show in greater details the derivation of this calculation and will highlight similar calculations for other observables for the sake of compactness. The $n$-th order contribution to the mean coordinate of the light flux reads:
\begin{equation}
\label{eq:mean_pos_1}
\begin{aligned}
    \langle \delta\bm{r}(t) \rangle^{(n)} &= (\mu_s c)^n e^{-\mu_s c t} \int d\bm{r} \left(\prod_{i=1}^n \int\limits_0^{t_{i+1}} dt_i\right) \left(\prod_{j=1}^{n} \int\limits_{\mathbb{S}^2} d\hat{\bm{s}}_j f(\hat{\bm{s}}_{j} \cdot \hat{\bm{s}}_{j-1})\right) \times\\
    &\times
    \bm{r} \delta^3(\bm{r} - c \sum\limits_{k=0}^n \hat{\bm{s}}_k (t_{k+1} - t_k)).
\end{aligned}
\end{equation}
The space integral can be carried out with help of the Dirac $\delta$-function at $\bm{r}=c \sum\limits_{k=0}^n \hat{\bm{s}}_k (t_{k+1} - t_k)$ yielding:
\begin{equation}
\label{eq:mean_pos_2}
\begin{aligned}
    \langle \delta\bm{r}(t) \rangle^{(n)} &= c (\mu_s c)^n e^{-\mu_s c t} \left(\prod_{i=1}^n \int\limits_0^{t_{i+1}} dt_i\right) \left(\prod_{j=1}^{n} \int\limits_{\mathbb{S}^2} d\hat{\bm{s}}_j f(\hat{\bm{s}}_{j} \cdot \hat{\bm{s}}_{j-1})\right) \times\\
    &\times
    \sum\limits_{k=0}^n \hat{\bm{s}}_k (t_{k+1} - t_k).
\end{aligned}
\end{equation}
The calculation of these integrals can be carried out exactly using results of~\cref{app:nested_time_integrals} for nested time integrals and of~\cref{app:moments_n_scattering_HG} for chained moments of the scattering function. Therefore,
\begin{equation}
\label{eq:mean_pos_3}
\begin{aligned}
    \langle \delta\bm{r}(t) \rangle^{(n)} &= c(\mu_sc)^{n}\sum\limits_{k=0}^{n}\bm{M}^{k} J_k\\
    & = c(\mu_sc)^{n}\sum\limits_{k=0}^{n}\frac{t^{n+1}}{(n+1)!}g^k\hat{\bm{s}}_0  = c(\mu_sc)^{n}\frac{t^{n+1}(1-g^{n+1})}{(1-g)(n+1)!}\hat{\bm{s}}_0.
\end{aligned}
\end{equation}
Therefore, the mean coordinate, summed over all scattering orders, reads:
\begin{equation}
    \langle \bm{r}(t) \rangle = \sum\limits^{\infty}_{n = 0} \langle \delta\bm{r}(t) \rangle^{(n)} =  c\sum\limits^{\infty}_{n = 0} (\mu_sc)^{n}e^{-\mu_sct}\frac{t^{n+1}(1-g^{n+1})}{(1-g)(n+1)!}\hat{\bm{s}}_0.
\label{eq:mean_pos_4}
\end{equation}
For convinience let us introduce the following notations:
\begin{equation}
\begin{aligned}
n_s       &= \mu_sct,\\
\mu_s'  &= \mu_s(1-g),\\
\mu_s'' &= \mu_s(1-g^2).
\end{aligned}
\end{equation}
In these notations~\cref{eq:mean_pos_4} reads:
\begin{equation}
    \langle \bm{r}(t) \rangle =  \sum\limits^{\infty}_{n = 0} e^{-n_s}\frac{n_s^{n+1}(1-g^{n+1})}{\mu_s'(n+1)!}\hat{\bm{s}}_0\\
    = \frac{1}{\mu_s'}(1-\exp(-\mu_s'ct)) \hat{\bm{s}}_0,
\end{equation}
where in the second line
\begin{equation}
    \sum\limits^{\infty}_{n = 0} \frac{n_s^{n+1}(1-g^{n+1})}{(n+1)!} = e^{n_s} - e^{gn_s},
\end{equation}
was used.

\subsection{Mean direction and velocity}
\label{app:moments_light_flux_mean_direction}
Let us calculate mean photon's direction and velocity. We begin with
\begin{equation}
\label{eq:mean_direction_1}
\begin{aligned}
\langle \delta\hat{\bm{s}}(t) \rangle^{(n)}
&= (\mu_sc)^n e^{-\mu_sct} \int d\bm{r} \left(\prod_{i=1}^n\int\limits_0^{t_{i+1}} dt_i\right)\left(\prod_{j=1}^{n}\int\limits_{\bm{S}^2}d\hat{\bm{s}}_{j} f(\hat{\bm{s}}_{j}\cdot\hat{\bm{s}}_{j-1})\right)\hat{\bm{s}}\times\\
&\times\delta^3(\bm{r} - c \sum\limits_{k=0}^n \hat{\bm{s}}_k (t_{k+1} - t_k))\\
&=(\mu_sc)^n e^{-\mu_sct} \left(\prod_{i=1}^n\int\limits_0^{t_{i+1}} dt_i\right)\left(\prod_{j=1}^{n}\int\limits_{\bm{S}^2}d\hat{\bm{s}}_{j} f(\hat{\bm{s}}_{j}\cdot\hat{\bm{s}}_{j-1})\right)\hat{\bm{s}}\\
&=(\mu_sc)^{n}e^{-\mu_sct}\bm{M}^n\frac{t^n}{n!}=(\mu_scg)^{n}e^{-\mu_sct}\frac{t^{n}}{n!}\hat{\bm{s}}_0,
\end{aligned}
\end{equation}
where in the third line the following equation was used:
\begin{equation}
    \left(\prod_{i=1}^n\int\limits_0^{t_{i+1}} dt_i\right) = \frac{t^n}{n!}
\label{eq:product_time_integrals}
\end{equation}
as well as~\cref{eq:moments_definition_chained_results}. Therefore,
\begin{equation}
    \langle \hat{\bm{s}}(t) \rangle = \sum\limits_{n=0}^{\infty}\langle \delta\hat{\bm{s}}(t) \rangle^{(n)} = e^{-\mu_sct}\sum\limits_{n=0}^{\infty}(\mu_scg)^{n}\frac{t^{n}}{n!}\hat{\bm{s}}_0 = \exp(-\mu_s'ct)\hat{\bm{s}}_0.
\end{equation}
Finally, since mean velocity is simply $\langle \bm{v}(t) \rangle  = \langle c\hat{\bm{s}}(t) \rangle$
one finds
\begin{equation}
    \langle \bm{v}(t) \rangle = c  \exp(-\mu_s'ct)\hat{\bm{s}}_0.
\end{equation}

\subsection{Directional Covariance Matrix}
\label{app:second_moments_light_flux_direction}
Following the approach in~\cref{eq:mean_direction_1}, we derive the expression for the average:
\begin{equation}
\begin{aligned}
    \langle \delta\left(\hat{\bm{s}}_i \hat{\bm{s}}_j\right) \rangle^{(n)} &= (\mu_s c)^n e^{-\mu_s c t} \left(\prod_{i=1}^n \int_0^{t_{i+1}} dt_i \right) \left(\prod_{j=1}^n \int_{\mathbb{S}^2} d\hat{\bm{s}}_j f(\hat{\bm{s}}_j \cdot \hat{\bm{s}}_{j-1}) \right) \hat{\bm{s}}_i \hat{\bm{s}}_j \\
    &=(\mu_s c)^n e^{-\mu_s c t} M^{nn}_{ij} \frac{t^n}{n!} \\
    &= \frac{(\mu_s c t)^n}{n!} e^{-\mu_s c t} \left( \frac{1-g^{2n}}{3} \delta_{ij} + g^{2n} (\hat{\bm{s}}_0)_i (\hat{\bm{s}}_0)_j \right),
\end{aligned}
\end{equation}
where we used~\cref{eq:moments_definition_chained,eq:product_time_integrals} in the second line and~\cref{eq:moments_definition_chained_results} in the third line. To avoid cluttering we omit indication of time dependence of the left-hand-side of
$\langle \delta\left(\hat{\bm{s}}_i \hat{\bm{s}}_j\right) \rangle^{(n)}$ here and in what follows.

The $ij$-element of the directional correlation matrix read:
\begin{equation}
\begin{aligned}
    \langle \hat{\bm{s}}_i \hat{\bm{s}}_j \rangle &= \sum_{n=0}^{\infty} \langle \delta\left(\hat{\bm{s}}_i \hat{\bm{s}}_j\right) \rangle^{(n)} \\
    &= \sum_{n=0}^{\infty} \frac{(\mu_s c t)^n}{n!} e^{-\mu_s c t} \left( \frac{1-g^{2n}}{3} \delta_{ij} + g^{2n} (\hat{\bm{s}}_0)_i (\hat{\bm{s}}_0)_j \right) \\
    &= \frac{1}{3} \delta_{ij} (1 - e^{-\mu_s'' c t}) + e^{-\mu_s'' c t} (\hat{\bm{s}}_0)_i (\hat{\bm{s}}_0)_j,
\end{aligned}
\end{equation}
where the following equation
\begin{equation}
    \sum_{n=0}^{\infty} \frac{(\mu_s c t)^n}{n!} e^{-\mu_s c t} g^{2n} = e^{-\mu_s c t (1-g^2)} = e^{-\mu_s'' c t}
\end{equation}
determines a new effective inverse scattering length $\mu_s'' = \mu_s (1-g^2)$.

The directional covariance matrix elements read:
\begin{equation}
\begin{aligned}
    D_{ij}(t)
    &= \langle (\hat{\bm{s}}_i-\langle \hat{\bm{s}}_i\rangle) (\hat{\bm{s}}_j-\langle \hat{\bm{s}}_j\rangle)\rangle = \langle \hat{\bm{s}}_i \hat{\bm{s}}_j\rangle - \langle \hat{\bm{s}}_i\rangle\langle \hat{\bm{s}}_j\rangle\\
    &= \frac{1}{3} \delta_{ij} (1 - e^{-\mu_s'' c t}) +  (\hat{\bm{s}}_0)_i (\hat{\bm{s}}_0)_j\left(e^{-\mu_s'' c t}-e^{-2\mu_s' c t}\right).
\end{aligned}
\end{equation}
Six independent matrix elements $D_{ij}$  fully characterize the dispersion of light flux direction. If a reader is interested in a single number instead, the following quantity might be helpful:
\begin{equation}
    D(t) = \text{Tr}(\langle \hat{\bm{s}}_i \hat{\bm{s}}_j \rangle - \langle \hat{\bm{s}}_i \rangle \langle \hat{\bm{s}}_j \rangle) = 1 - e^{-2\mu_s' ct}.
\end{equation}
One can observe that while
\subsection{Direction-Spatial Covariance Matrix}
\label{app:second_moments_light_flux_direction_space}
Here we examine the following correlation matrix element $\langle \bm{r}_i \hat{\bm{s}}_j \rangle$. As above we begin with its definition for the $n$-th scattering order:
\begin{equation}
\begin{aligned}
\langle \delta\left(\bm{r}_i \hat{\bm{s}}_j\right) \rangle^{(n)} &= (\mu_s c)^n e^{-\mu_s c t} \int d\bm{r} \left(\prod_{i=1}^n \int_0^{t_{i+1}} dt_i\right) \left(\prod_{j=1}^n \int_{\bm{S}^2} d\hat{\bm{s}}_j f(\hat{\bm{s}}_{j} \cdot \hat{\bm{s}}_{j-1})\right) \times \\
&\times  \bm{r}_i (\hat{\bm{s}})_j \delta^3 \left(\bm{r} - c \sum_{k=0}^n \hat{\bm{s}}_k (t_{k+1} - t_k)\right)\\
&= c (\mu_s c)^n e^{-\mu_s c t} \left(\prod_{i=1}^n \int_0^{t_{i+1}} dt_i\right) \left(\prod_{j=1}^n \int_{\bm{S}^2} d\hat{\bm{s}}_j f(\hat{\bm{s}}_{j} \cdot \hat{\bm{s}}_{j-1})\right) \times \\
&\times  \sum_{k=0}^n (\hat{\bm{s}}_k)_i (t_{k+1} - t_k) (\hat{\bm{s}})_j,
\end{aligned}
\end{equation}
where the Dirac $\delta$-function was used to carry out the space integral at last two lines. Angular and time integrals can be carried out using results of~\cref{app:moments_HG,app:moments_n_scattering_HG}, in particular,~\cref{eq:moments_definition_chained,disp:HG}, yielding:
\begin{equation}
\begin{aligned}
\langle \delta\left(\bm{r}_i \hat{\bm{s}}_j\right) \rangle^{(n)}
&= c (\mu_s c)^n e^{-\mu_s c t} \sum_{k=0}^n M^{kn}_{ij} J_k \\
&= c (\mu_s c)^n e^{-\mu_s c t} \sum_{k=0}^n \frac{t^{n+1}}{(n+1)!} g^{n-k} \left(\frac{1 - g^{2k}}{3} \delta_{ij} + g^{2k} (\hat{\bm{s}}_0)_i (\hat{\bm{s}}_0)_j\right) \\
&= \frac{(\mu_s c t)^{n+1}}{\mu_s (n+1)!} e^{-\mu_s c t} g^n \sum_{k=0}^n \left(\frac{g^{-k} - g^k}{3} \delta_{ij} + g^k (\hat{\bm{s}}_0)_i (\hat{\bm{s}}_0)_j\right).
\end{aligned}
\end{equation}
Taking into account that
\begin{equation}
    \sum_{k=0}^n (g^{-k} - g^k) = \frac{g^{-n}(1 - g^n)(1 - g^{n+1})}{1 - g}
\end{equation}
and
\begin{equation}
    \sum_{k=0}^n g^k = \frac{1 - g^{n+1}}{1 - g},
\end{equation}
one can obtain
\begin{equation}
\begin{aligned}
\langle \delta\left(\bm{r}_i \hat{\bm{s}}_j\right) \rangle^{(n)} &= \frac{(\mu_s c t)^{n+1}}{\mu_s' (n+1)!} e^{-\mu_s c t} (1 - g^{n+1}) \left(\frac{1 - g^n}{3} \delta_{ij} + g^n (\hat{\bm{s}}_0)_i (\hat{\bm{s}}_0)_j\right).
\end{aligned}
\end{equation}
The correlation due to all scatterings is obtained summing over all $n$:
\begin{equation}
\begin{aligned}
\langle \bm{r}_i \hat{\bm{s}}_j \rangle &= \sum_{n=0}^\infty \langle \delta\left(\bm{r}_i \hat{\bm{s}}_j\right) \rangle^{(n)} \\
&= \sum_{n=0}^\infty \frac{(\mu_s c t)^{n+1}}{\mu_s' (n+1)!} e^{-\mu_s c t} (1 - g^{n+1}) \left(\frac{1 - g^n}{3} \delta_{ij} + g^n (\hat{\bm{s}}_0)_i (\hat{\bm{s}}_0)_j\right).
\end{aligned}
\end{equation}
All three terms can be summed up exactly, using:
\begin{equation}
\begin{aligned}
    e^{-n_s} \sum_{n=0}^\infty \frac{n_s^{n+1}}{(n+1)!} &= \frac{1}{g} (1 - e^{-n_s}),\\
    e^{-n_s} \sum_{n=0}^\infty \frac{n_s^{n+1} g^n}{(n+1)!} &= \frac{1}{g} (e^{-(1-g)n_s} - e^{-n_s}),\\
    e^{-n_s} \sum_{n=0}^\infty \frac{n_s^{n+1} g^{2n+1}}{(n+1)!} &= \frac{1}{g} (e^{-(1-g^2)n_s} - e^{-n_s}).
\end{aligned}
\end{equation}
Therefore, finaly
\begin{equation}
\begin{aligned}
    \langle \bm{r}_i \hat{\bm{s}}_j \rangle &= \frac{1}{\mu_s' g} \left[\frac{1}{3} (e^{-(1-g^2)n_s} + g - (1 + g) e^{-(1-g)n_s}) \delta_{ij} \right. \\
    &+ \left. (e^{-(1-g)n_s} - e^{-(1-g^2)n_s}) (\hat{\bm{s}}_0)_i (\hat{\bm{s}}_0)_j \right],
\end{aligned}
\label{eq:direction_spatial_covariance}
\end{equation}
where $n_s=\mu_s ct$.

Summing up~\cref{eq:direction_spatial_covariance} over the $i=j$ indices yields:
\begin{equation}
    \langle \bm{r} \cdot \hat{\bm{s}} \rangle = \sum_{i=1}^3 \sum_{j=1}^3 \langle \bm{r}_i \hat{\bm{s}}_j \rangle \delta_{ij} = \frac{1}{\mu_s'} (1 - e^{-(1-g^2)n_s}).
\end{equation}

\subsection{Spatial Covariance Matrix}
\label{app:second_moments_light_flux_space}
The spatial correlation matrix is determined by:
\begin{equation}
\begin{aligned}
\langle \delta\left(\bm{r}_i \bm{r}_j\right) \rangle^{(n)}
&= (\mu_s c)^n e^{-\mu_s c t} \int d\bm{r} \left(\prod_{i=1}^n \int_0^{t_{i+1}} dt_i\right) \left(\prod_{j=1}^n \int_{\bm{S}^2} d\hat{\bm{s}}_j f(\hat{\bm{s}}_{j} \cdot \hat{\bm{s}}_{j-1})\right) \times \\
&\times \bm{r}_i \bm{r}_j \delta^3\left(\bm{r} - c \sum_{k=0}^n \hat{\bm{s}}_k (t_{k+1} - t_k)\right)\\
&= c^2(\mu_sc)^n e^{-\mu_sct} \left(\prod_{i=1}^n\int\limits_0^{t_{i+1}} dt_i\right)\left(\prod_{j=1}^{n}\int\limits_{\bm{S}^2}d\hat{\bm{s}}_{j} f(\hat{\bm{s}}_{j}\cdot\hat{\bm{s}}_{j-1})\right) \times\\
&\times
\sum\limits_{k=0}^{n}(\hat{\bm{s}}_k)_i(t_{k+1}-t_k)\sum\limits_{l=0}^{n}(\hat{\bm{s}}_l)_j(t_{l+1}-t_l)\\
&=c^2(\mu_sc)^n e^{-\mu_sct} \left(\prod_{i=1}^n\int\limits_0^{t_{i+1}} dt_i\right)\left(\prod_{j=1}^{n}\int\limits_{\bm{S}^2}d\hat{\bm{s}}_{j} f(\hat{\bm{s}}_{j}\cdot\hat{\bm{s}}_{j-1})\right) \times\\
&\times\left(
2\sum\limits_{k<l}(\hat{\bm{s}}_k)_i(\hat{\bm{s}}_l)_j(t_{k+1}-t_k)(t_{l+1}-t_l)+\sum\limits_{k=0}^{n}(\hat{\bm{s}}_k)^2_i(t_{k+1}-t_k)^2\right)
\end{aligned}
\end{equation}
where~\cref{eq:moments_definition_chained_results} and ~\cref{eq:double_momentum_for_time} were used, also accounting for an obvious relationship for any matrix $A$:
\begin{equation}
    \sum_{k=0}^n \sum_{l=0}^n A_{kl} = 2 \sum_{k<l} A_{kl} + \sum_{k=l} A_{kk}.
\end{equation}
Expanding this relationship, one can get:
\begin{equation}
    \begin{aligned}
    \langle \delta\left(\bm{r}_i \bm{r}_j\right) \rangle^{(n)}&=c^2(\mu_sc)^n e^{-\mu_sct}\left(
2\sum\limits_{k<l}M_{ij}^{kl}J_{kl}+\sum\limits_{k=0}^{n}M_{ii}^{kk}J_{kk}\right)\\
&=c^2(\mu_sc)^n e^{-\mu_sct}\left(
2\sum\limits_{k<l}M_{ij}^{kl}J_{00}+2\sum\limits_{k=0}^{n}M_{ii}^{kk}J_{00}\right)\\
& = 2c^2(\mu_sc)^n e^{-\mu_sct}\frac{t^{n+2}}{(n+2)!}\left(\sum\limits_{k=0}^{n}\sum\limits_{l=k+1}^{n} M_{ij}^{kl} +\sum\limits_{k=0}^{n}M_{ii}^{kl}\right)\\
&=2c^2(\mu_sc)^n e^{-\mu_sct}\frac{t^{n+2}}{(n+2)!}\left(\sum\limits_{k=0}^{n}\frac{1 - g^{2k}}{3} \delta_{ij} + g^{2k} (\hat{\bm{s}}_0)_i (\hat{\bm{s}}_0)_j\right)\times \\
&\times \left(1+ \sum\limits_{l=k+1}^{n}g^{l-k}\right).
    \end{aligned}
\end{equation}
Taking into account, that
\begin{equation}
    \begin{aligned}
        1+\sum\limits_{l=k+1}^{n}g^{l-k} = 1+\frac{g(1-g^{n-k})}{1-g} = \frac{1-g^{n-k}}{1-g},
    \end{aligned}
\end{equation}
one can get the following:
\begin{equation}
    \begin{aligned}
\langle \delta\left(\bm{r}_i \bm{r}_j\right) \rangle^{(n)}&=2c^2(\mu_sc)^n e^{-\mu_sct}\frac{t^{n+2}}{(n+2)!}\left(\sum\limits_{k=0}^{n}\frac{1 - g^{2k}}{3} \delta_{ij} + g^{2k} (\hat{\bm{s}}_0)_i (\hat{\bm{s}}_0)_j\right)\times \\
&\times \left(\frac{1-g^{n-k}}{1-g}\right).
    \end{aligned}
\end{equation}
Finally,
\begin{equation}
\begin{aligned}
\langle \bm{r}_i \bm{r}_j \rangle &= \sum\limits_{n=0}^\infty \langle \delta\left(\bm{r}_i \bm{r}_j\right) \rangle^{(n)} =\frac{2}{(\mu_s')^2(1+g)g}\times \\
&\times\left[- \frac{1}{3} \left(g(g+2+\mu_sct(g^2-1)) -(1+g)^2e^{-\mu_s'ct} + e^{-\mu_s''ct}\right)\delta_{ij}\right. \\
&\left.+ \left(g-(1+g)e^{-\mu_s'ct} + e^{-\mu_s''ct}\right)\hat{\bm{s}}_0^i\hat{\bm{s}}_0^j\right].
\end{aligned}
\end{equation}

\section{Derivation of Time-Integrated Flux within a Sphere}
\label{app:derivation_time_integrated_flux}
Here we consider the following quantity
\begin{equation}
\begin{aligned}
G(r,\bm{s}) &\equiv \int\limits_0^\infty dt\int\limits_{\bm{S}^2} d\hat{\bm{r}} L(\bm{r},t,\hat{\bm{s}}),
\end{aligned}
\label{app:spherical_fluence1}
\end{equation}
where $\hat{\bm{r}}$ is the unit vector along $\bm{r}$. Substituting~\cref{eq:L_final_solution} for $L$ into~\cref{app:spherical_fluence1} yields:

To carry out the integrals, the following calculation might be helpful:
\begin{equation}
\begin{aligned}
    \int_0^\infty dt &\int_{\bm{S}^2} d\hat{\bm{r}} e^{-\mu_t ct} \delta^3(\bm{r} - c \hat{\bm{s}}_0 t)\\
    &= \int_0^\infty dt \int_{\bm{S}^2} d\hat{\bm{r}} e^{-\mu_t ct} \frac{1}{cr^2} \delta^2(\hat{\bm{r}} - \hat{\bm{s}}_0) \delta(t - r/c)= \frac{1}{cr^2} e^{-\mu_t r}.
\end{aligned}
\end{equation}
Therefore,
\begin{equation}
\begin{aligned}
\int_0^\infty dt &\int_{\bm{S}^2} d\hat{\bm{r}} e^{-\mu_t ct} (\mu_s ct)^n \delta^3(\bm{r} - ct \bm{s}^{(n)})\\
 &= \int_0^\infty dt \int_{\bm{S}^2} d\hat{\bm{r}} e^{-\mu_t ct} (\mu_s ct)^n \frac{1}{cr^2} \delta^2 \left(\hat{\bm{r}} - \frac{\bm{s}^{(n)}}{|\bm{s}^{(n)}|}\right) \delta \left(t - \frac{r}{c |\bm{s}^{(n)}|}\right) \\
 &= \frac{1}{cr^2 |\bm{s}^{(n)}|} e^{-\mu_t r/|\bm{s}^{(n)}|} \left(\frac{\mu_s r}{|\bm{s}^{(n)}|}\right)^n.
\end{aligned}
\end{equation}
Taking these integrals into account, we can integrate the~\cref{app:spherical_fluence1}
\begin{equation}
\begin{aligned}
G(r, \bm{s}) &\equiv \int_0^\infty dt \int_{\bm{S}^2} d\hat{\bm{r}} L(\bm{r}, t, \hat{\bm{s}}) = \frac{1}{r^2} e^{-\mu_t r} \delta^2(\hat{\bm{s}} - \hat{\bm{s}}_0) \\
&+ \frac{1}{r^2} \sum_{n=1}^\infty \left(\prod_{i=1}^n \int_0^1 d\xi_i (1-\xi_i)^{i-1}\right) \left(\prod_{j=1}^{n-1} \int_{\bm{S}^2} d\hat{\bm{s}}_j f_g(\hat{\bm{s}}_{j+1} \cdot \hat{\bm{s}}_j)\right) \\
&\times f_g(\hat{\bm{s}}_1 \cdot \hat{\bm{s}}_0) \left(\frac{\mu_s r}{|\bm{s}^{(n)}|}\right)^n \frac{e^{-\mu_t r/|\bm{s}^{(n)}|}}{|\bm{s}^{(n)}|}
\end{aligned}
\end{equation}

\section{Probabilistic Derivation of the Solution}
\label{app:probabilistic_derivaion}
We shall demonstrate that~\cref{eq:Ln_general_unidir} can be derived purely from principles of probability theory, without recourse to the RTE. This approach not only clarifies the physical underpinnings of the various terms but also reinforces the validity of the derived result. Consider a sequence of $n$ independent scattering events occurring within the time interval $(t_0, t)$. These events are randomly distributed over times $t_1, t_2, \ldots, t_n$ and directions $\hat{\bm{s}}_1, \hat{\bm{s}}_2, \ldots, \hat{\bm{s}}_n$. The probability density function (PDF) of an exponential distribution is defined as:
\[ p(t;\lambda) = \lambda e^{-\lambda t}. \]

We postulate the following chronological ordering of scattering events:
\begin{equation}
t_0 \le t_1 \le t_2 \le \ldots \le t_n \le t.
\label{eq:times_ordering1}
\end{equation}

Under this assumption, the joint PDF becomes:
\begin{equation}
\begin{aligned}
p(t_1,t_2,\ldots, t_n;\lambda)
&= \lambda e^{-\lambda (t_1-t_0)} \cdot \lambda e^{-\lambda (t_2-t_1)} \ldots \lambda e^{-\lambda (t_n-t_{n-1})} \cdot e^{-\lambda (t-t_{n})} \\
&= \lambda^n e^{-\lambda t},
\end{aligned}
\end{equation}
where each term $\lambda e^{-\lambda (t_i-t_{i-1})}$ represents the PDF for the time interval $(t_{i-1}, t_i)$, and the final term in the first line signifies the probability of no scattering event occurring in the interval $(t_n, t)$.


We have to sum over all possible scattering times and directions, imposing the requirement that every photon's trajectory, starting at point $\bm{r}_0$ at time $t_0$ and along an initial direction $\hat{\bm{s}}_0$, should end up at point $\bm{r}$ exactly at time $t$ with direction $\hat{\bm{s}}$. This is precisely accounted for by incorporating Dirac delta functions:
\begin{equation}
\delta^3(\bm{r}_0 + c(t_1 - t_0)\hat{\bm{s}}_0 + c(t_2 - t_1)\hat{\bm{s}}_1 + \dots + c(t - t_n)\hat{\bm{s}}_n - \bm{r}) \cdot \delta^2(\hat{\bm{s}}_n - \hat{\bm{s}}),
\end{equation}
where the first $\delta$-function ensures that the photon's path concludes at the specified position $\bm{r}$ at time $t$, summing over all possible combinations of scattering events. The second $\delta$-function confirms that the final scattering aligns with the necessary direction
$\hat{\bm{s}}$, completing the photon's journey as required.

For assumed ordering in~\cref{eq:times_ordering1} the allowed intervals are given by
\begin{equation}
t_1\in (t_0,t_2), \quad t_2\in (t_0,t_3), \quad  \dots \quad  t_n\in (t_0,t)
\end{equation}
The probability $P(\bm{r},t,\hat{\bm{s}})$ to observe a photon at point $\bm{r}$ exactly at time $t$ with direction $\hat{\bm{s}}$ reads:
\begin{equation}
\begin{aligned}
P(\bm{r},t,\hat{\bm{s}})
&= (\mu_s c)^n e^{-\mu_s c t}\int\limits_{t_0}^t dt_n \dots  \int\limits_{t_0}^{t_3} dt_2\int\limits_{t_0}^{t_2} dt_1
\int\limits_{\bm{S}^2}d\hat{\bm{s}}_{1}
\int\limits_{\bm{S}^2}d\hat{\bm{s}}_{2}\dots
\int\limits_{\bm{S}^2}d\hat{\bm{s}}_{n}\\
&\times f(\hat{\bm{s}}_{0}\cdot\hat{\bm{s}}_{1})
f(\hat{\bm{s}}_{1}\cdot\hat{\bm{s}}_{2})\dots
f(\hat{\bm{s}}_{n-1}\cdot\hat{\bm{s}}_{n})\\
&\times \delta^3(\bm{r}_0+c(t_1-t_0)\hat{\bm{s}}_0+c(t_2-t_1)\hat{\bm{s}}_1+\dots c(t-t_{n})\hat{\bm{s}}_{n}-\bm{r})\\
&\times \delta^2(\hat{\bm{s}}_{n}-\hat{\bm{s}}) e^{-\mu_a ct},
\end{aligned}
\label{eq:probability_n_order}
\end{equation}
where the last factor corresponds to the probability of no absorption.
Multiplying the probability $P(\bm{r},t,\hat{\bm{s}})$ by speed of light in the medium $c$ one gets the flux in~\cref{eq:deltan_0}.

\section{Observable Signal}
\label{sec:observables}
In~\cref{sec:expansion_series_unidirectional}, we have discussed the source function $L$, which is a singular function. However, in the context of physical observables, such as the number ($\Delta N$) of detected photons, it is essential that these observables possess finite measures everywhere.

As an illustrative example, let's consider a detector characterized by a photo-detection efficiency function $\varepsilon(\hat{\bm{n}},\hat{\bm{s}})$, where $\hat{\bm{n}}$ is the normal to a point $\bm{r}$ on the detector's surface, and $\bm{\hat{s}}$ is the unit vector of a photon's direction.

\begin{figure}[!htb]
\begin{center}
\includegraphics[width=0.5\linewidth]{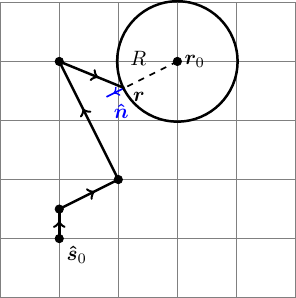}
\caption{An example of a photon's trajectory with an initial direction $\bm{\hat{s}}_0$, originating from the origin and intersecting the spherical detector at point $\bm{r}$. The sphere's center is displaced from the origin by $\bm{r}_0$. $\bm{\hat{n}}$ is the unit vector orthogonal to the surface at point $\bm{r}$.}
\label{fig:detector}
\end{center}
\end{figure}

For the purpose of this discussion, we'll focus on a spherical detector, similar to those used in Neutrino Telescopes. This detector has its center displaced from the light source by $\bm{r}_0$, and its radius is denoted as $R$ (see~\cref{fig:detector}).

A point on the spherical detector's surface can be expressed as $\bm{r} = \bm{r}_0+\bm{\hat{n}}R$, where $\bm{\hat{n}}$ is the appropriate unit vector as shown in~\cref{fig:detector}.

To estimate the number of detected photons within a time interval $(\tau_i,\tau_{i+1})$, we use the following integral:
\begin{equation}
\begin{aligned}
    \Delta N_i &= R^2\int\limits_{\tau_i}^{\tau_{i+1}}dt\int\limits_{\mathbb{S}^2}d\bm{\hat{n}}\int\limits_{\mathbb{S}^2}d\bm{\hat{s}} L(\bm{r},t,\bm{\hat{s}})\varepsilon(\bm{\hat{n}},\bm{\hat{s}})H(-\bm{\hat{n}}\cdot\bm{\hat{s}}),
\end{aligned}
\end{equation}
where the Heaviside function $H(-\bm{\hat{n}}\cdot\bm{\hat{s}})$ ensures that the photon reaches the detector from the outside (i.e., $\bm{\hat{n}}\cdot\bm{\hat{s}}$ must be negative).

The integration over time and surface variables ($t$ and $\bm{\hat{n}}$) can be readily performed using the $\delta^3$-function.

The expected signals for no-scattering ($\Delta N^{(0)}_i$) and for $n$ scatterings ($\Delta N^{(n)}_i$) in the time interval $(\tau_i,\tau_{i+1})$ are given by:
\begin{equation}
\label{eq:DeltaN}
    \begin{aligned}
    \Delta N^{(0)}_i  &= e^{-\mu_a ct^*}P_0(\mu_s ct^{*})\varepsilon(\bm{\hat{n}}^*,\bm{\hat{s}}_0)H_{i}\frac{R}{\sqrt{D}},\\
    \Delta N^{(n)}_i  &=
    \prod_{k=1}^{n}\left(\int\limits_{0}^{1}d\xi_k (1-\xi_k)^{k-1}\int\limits_{\mathbb{S}^2}d\bm{\hat{s}}_k g(\bm{\hat{s}}_{k},\bm{\hat{s}}_{k-1})\right)\\
    &\times e^{-\mu_a ct^*}P_n(\mu_s ct^*)\varepsilon(\bm{\hat{n}}^*,\hat{\bm{s}}) \frac{n!R}{\sqrt{D^{(n)}}}H_{i},
    \end{aligned}
\end{equation}
where
\begin{equation}
\label{eq:intersection1}
\begin{aligned}
    \bm{\hat{n}}^* & = \frac{ct^{*}\bm{s}^{(n)} - \bm{r}_0}{R},\\
    D^{(n)} & = (\bm{r}_0\cdot\bm{s}^{(n)})^2 - (\bm{s}^{(n)}\cdot\bm{s}^{(n)})(\bm{r}_0^2-R^2),\\
    ct^* &= \frac{1}{\bm{s}^{(n)}\cdot\bm{s}^{(n)}}\left(\bm{r}_0\cdot\bm{s}^{(n)}- \sqrt{D^{(n)}}\right).
\end{aligned}
\end{equation}
Here, $D^{(n)}$ is the discriminant of the quadratic equation for the intersection of the line and the sphere,  $t^*$ is the corresponding solution representing the minimum intersection time, and $\bm{s}^{(n)}$ is given by~\cref{eq:s_n}.

The factor $H_{i}$ is defined as:
\begin{equation}
\label{eq:intersection2}
    H_{i} \equiv H(t^* - \tau_i)H(\tau_{i+1}-t^*)H(-\bm{\hat{n}}^*\cdot\hat{\bm{s}})H(D^{(n)}),
\end{equation}
ensuring that the intersection time lies within the selected time interval $\tau_i\le t^* \le \tau_{i+1}$, the photon reaches the detector surface from the outside ($\bm{\hat{n}}^*\cdot\hat{\bm{s}}<0$), and the intersection is physically possible ($D^{(n)}\ge 0$).

For the zero-order expectation ($\Delta N^{(0)}_i$), one should replace $\bm{s}^{(n)}\to \hat{\bm{s}}_0$ and $\bm{\hat{s}}_n\to \hat{\bm{s}}_0$ in~\cref{eq:intersection1,eq:intersection2}.

\Cref{eq:DeltaN,eq:intersection1,eq:intersection2} provide a comprehensive solution for the expected observable signal.

\section{Numerical Analysis of Contributions from First Four Scatterings}
\label{sec:test_orders}
\begin{figure}[!h]
    \centering
    \begin{tabular}{cc}
    \includegraphics[width=0.5\textwidth]{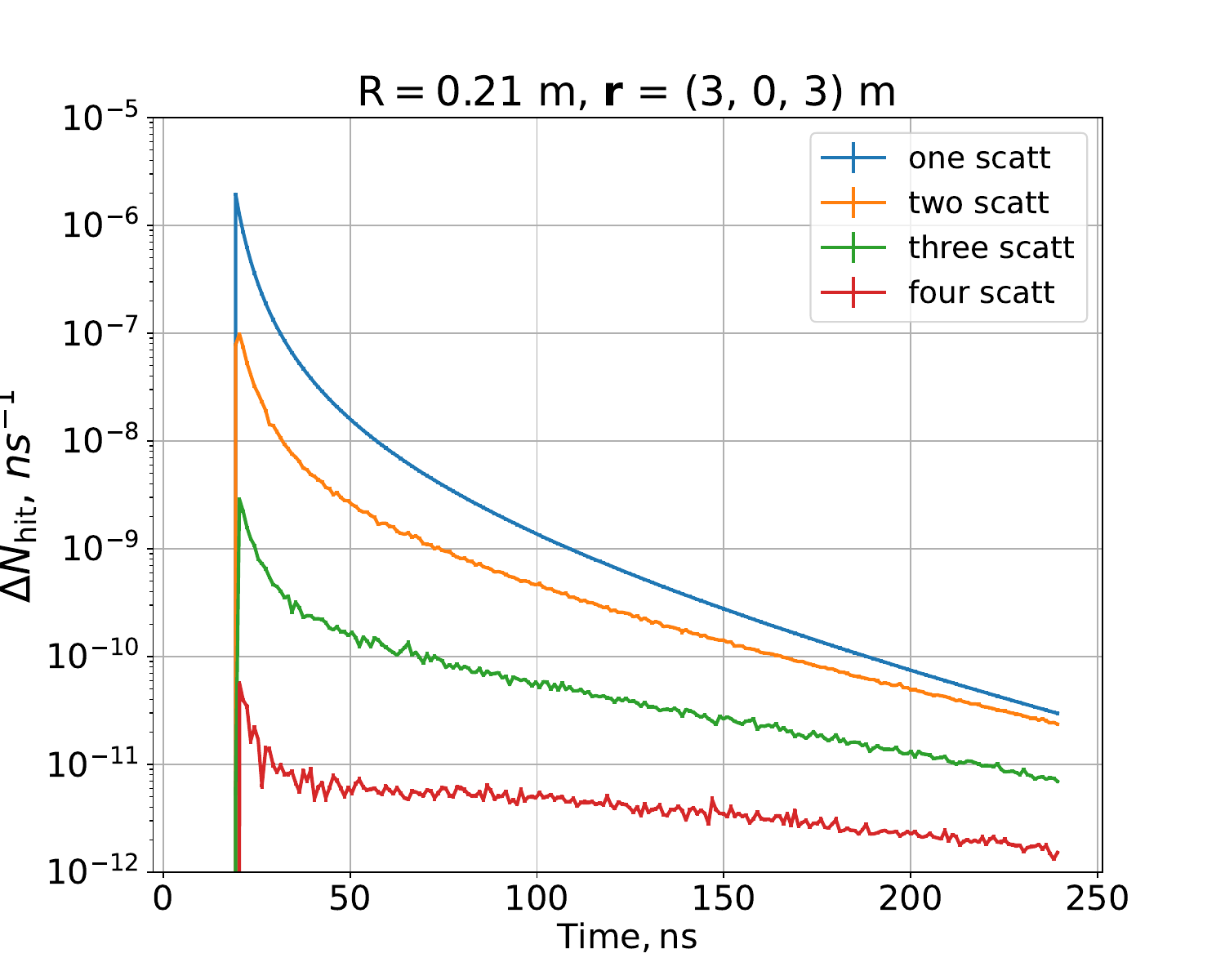}&
    \includegraphics[width=0.5\textwidth]{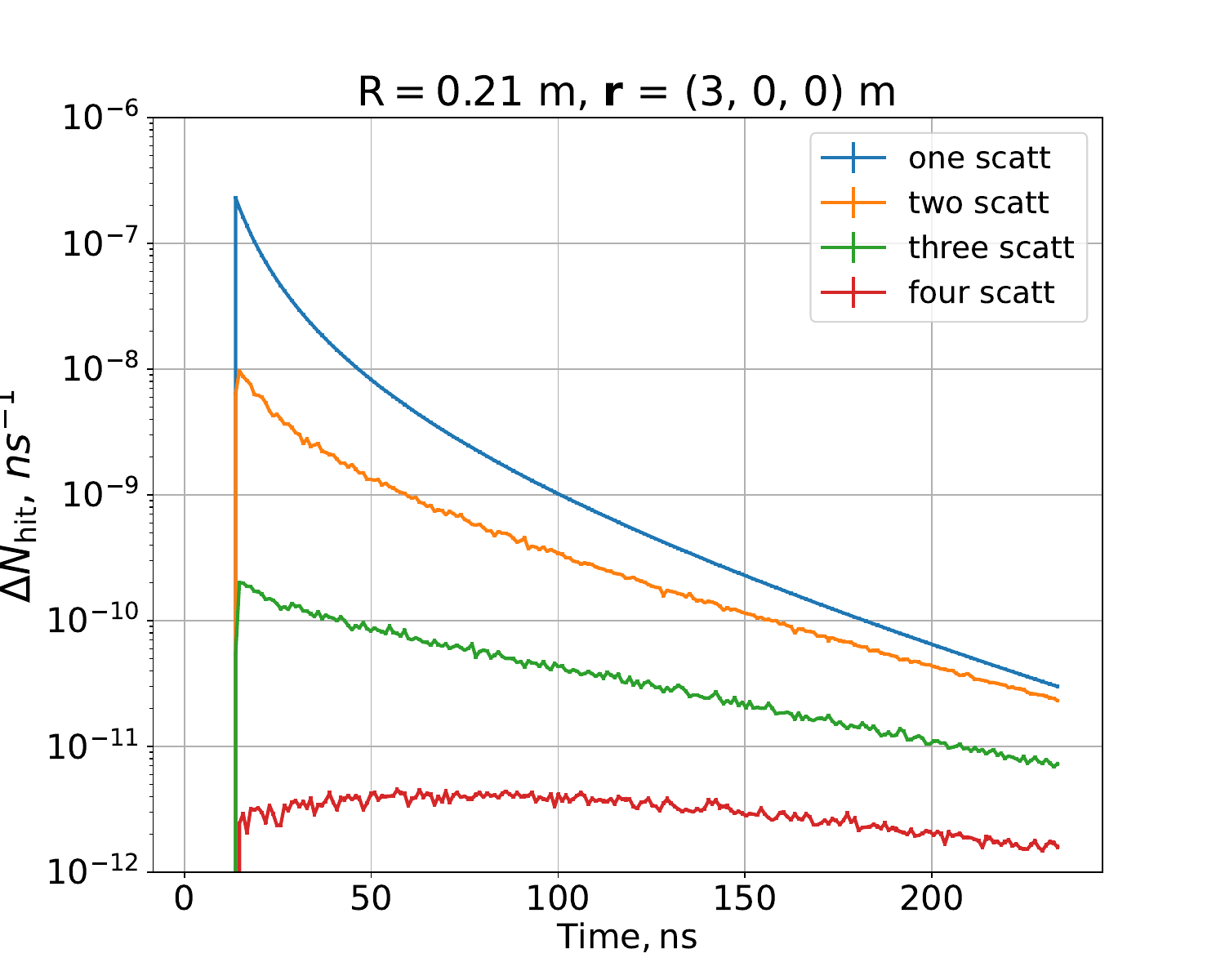}\\
    \includegraphics[width=0.5\textwidth]{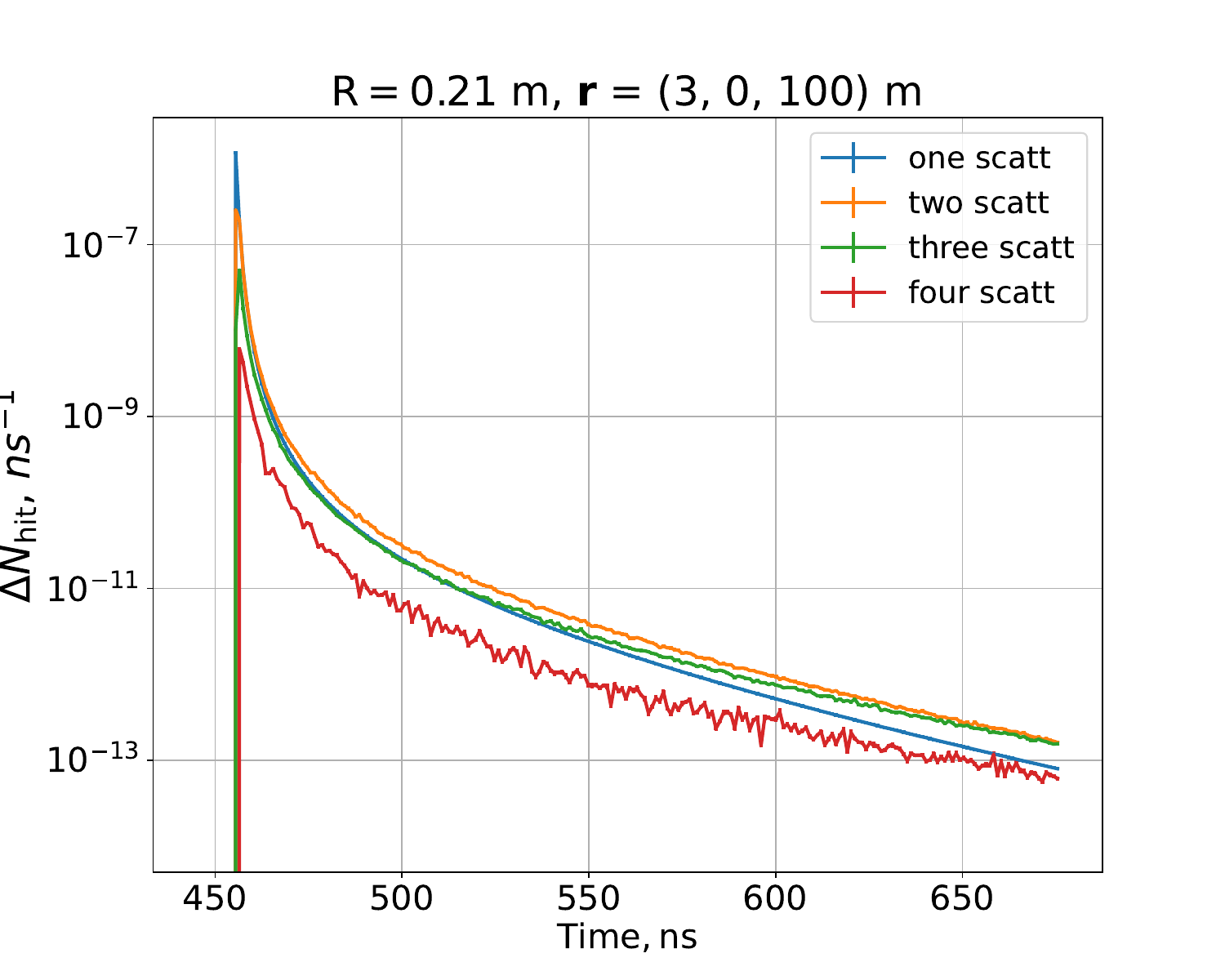}&
    \includegraphics[width=0.5\textwidth]{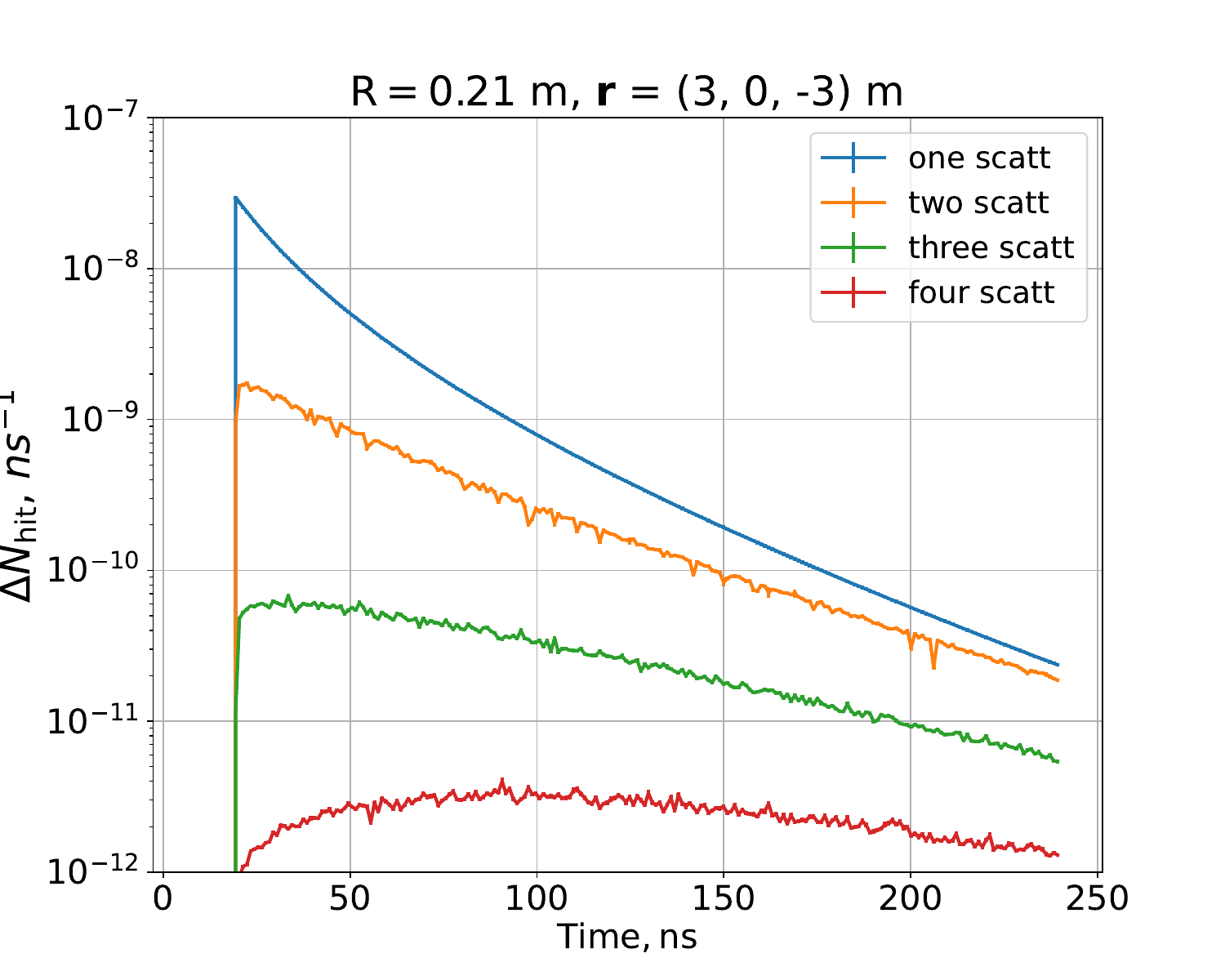}
    \end{tabular}
    \caption{Expected number of detected photons $\Delta N_\text{hit}$ integrated in one ns time bins obtained by our solution of the radiative transfer equation due to first four scatterings for four test points: $\bm{r} = (3,0,3)$ m (left, up), $\bm{r} = (3,0,0)$ m (right, up), $\bm{r} = (3,0,100)$ m (left, bottom), $\bm{r} = (3,0,-3)$ m (right, bottom). Note: The presence of noise in these distributions is a result of the Monte Carlo method used for VEGAS multi-dimensional integral calculations.}
    \label{fig:RTE_orders}
\end{figure}
\Cref{fig:RTE_orders} displays the expected number of detected photons $\Delta N_\text{hit}$ integrated in one ns time bins. These results were obtained by our solution of the radiative transfer equation, considering first four scatterings for four test points: $\bm{r} = (3,0,3)$ m, $\bm{r} = (3,0,0)$ m, $\bm{r} = (3,0,100)$ m, $\bm{r} = (3,0,-3)$ m and considering the Henyey-Greenstein scattering phase function with $g=0.9$ and assuming
$\mu_a^{-1} = 20.9$ m and $\mu_s^{-1} = 69.26$ m. The following assessments could be of importance.

(i) Notably, side and backward test points require a greater number of scatterings to accurately describe the expected signal, which is expected due to their locations. The contributions to the signal at these points increase more rapidly compared to the forward test point, as demonstrated in the left-bottom panel of~\cref{fig:RTE_orders}.

(ii) Higher-order scatterings ($n_\text{scatt}\ge 2$) match the contribution of single scattering approximately two hundred nanoseconds after the arrival of the first light.
\end{appendices}
\bibliographystyle{elsarticle-num}

\begin{thebibliography}{10}
\expandafter\ifx\csname url\endcsname\relax
  \def\url#1{\texttt{#1}}\fi
\expandafter\ifx\csname urlprefix\endcsname\relax\def\urlprefix{URL }\fi
\expandafter\ifx\csname href\endcsname\relax
  \def\href#1#2{#2} \def\path#1{#1}\fi

\bibitem{doi:10.1118/1.595361}
B.~C. Wilson, G.~Adam,
  \href{https://aapm.onlinelibrary.wiley.com/doi/abs/10.1118/1.595361}{A monte
  carlo model for the absorption and flux distributions of light in tissue},
  Medical Physics 10~(6) (1983) 824--830.
\newblock \href
  {http://arxiv.org/abs/https://aapm.onlinelibrary.wiley.com/doi/pdf/10.1118/1.595361}
  {\path{arXiv:https://aapm.onlinelibrary.wiley.com/doi/pdf/10.1118/1.595361}},
  \href {https://doi.org/10.1118/1.595361} {\path{doi:10.1118/1.595361}}.
\newline\urlprefix\url{https://aapm.onlinelibrary.wiley.com/doi/abs/10.1118/1.595361}

\bibitem{PhysRevLett.97.018104}
A.~Kienle, R.~Hibst,
  \href{https://link.aps.org/doi/10.1103/PhysRevLett.97.018104}{Light guiding
  in biological tissue due to scattering}, Phys. Rev. Lett. 97 (2006) 018104.
\newblock \href {https://doi.org/10.1103/PhysRevLett.97.018104}
  {\path{doi:10.1103/PhysRevLett.97.018104}}.
\newline\urlprefix\url{https://link.aps.org/doi/10.1103/PhysRevLett.97.018104}

\bibitem{Martelli_book}
F.~Martelli, S.~D. Bianco, A.~Ismaelli, G.~Zaccanti, Light Propagation Through
  Biological Tissue, SPIE Press, Bellingham,WA, 2010.

\bibitem{ALLISON2016186}
J.~Allison, et~al.,
  \href{http://www.sciencedirect.com/science/article/pii/S0168900216306957}{Recent
  developments in geant4}, Nuclear Instruments and Methods in Physics Research
  Section A: Accelerators, Spectrometers, Detectors and Associated Equipment
  835 (2016) 186 -- 225.
\newblock \href {https://doi.org/10.1016/j.nima.2016.06.125}
  {\path{doi:10.1016/j.nima.2016.06.125}}.
\newline\urlprefix\url{http://www.sciencedirect.com/science/article/pii/S0168900216306957}

\bibitem{Blyth:2021gam}
S.~Blyth, {Integration of JUNO simulation framework with Opticks: GPU
  accelerated optical propagation via NVIDIA\textregistered{}
  OptiX\texttrademark{}}, EPJ Web Conf. 251 (2021) 03009.
\newblock \href {https://doi.org/10.1051/epjconf/202125103009}
  {\path{doi:10.1051/epjconf/202125103009}}.

\bibitem{PhysRevLett.113.101101}
I.~C. Aartsen,
  \href{https://link.aps.org/doi/10.1103/PhysRevLett.113.101101}{Observation of
  high-energy astrophysical neutrinos in three years of icecube data}, Phys.
  Rev. Lett. 113 (2014) 101101.
\newblock \href {https://doi.org/10.1103/PhysRevLett.113.101101}
  {\path{doi:10.1103/PhysRevLett.113.101101}}.
\newline\urlprefix\url{https://link.aps.org/doi/10.1103/PhysRevLett.113.101101}

\bibitem{Baikal-GVD:2021zsq}
I.~Belolaptikov, et~al., {Neutrino Telescope in Lake Baikal: Present and
  Nearest Future}, PoS ICRC2021 (2021) 002.
\newblock \href {http://arxiv.org/abs/2109.14344} {\path{arXiv:2109.14344}},
  \href {https://doi.org/10.22323/1.395.0002} {\path{doi:10.22323/1.395.0002}}.

\bibitem{KM3Net:2016zxf}
S.~Adrian-Martinez, et~al., {Letter of intent for KM3NeT 2.0}, J. Phys. G
  43~(8) (2016) 084001.
\newblock \href {http://arxiv.org/abs/1601.07459} {\path{arXiv:1601.07459}},
  \href {https://doi.org/10.1088/0954-3899/43/8/084001}
  {\path{doi:10.1088/0954-3899/43/8/084001}}.

\bibitem{ANTARES:2011hfw}
M.~Ageron, et~al., {ANTARES: the first undersea neutrino telescope}, Nucl.
  Instrum. Meth. A 656 (2011) 11--38.
\newblock \href {http://arxiv.org/abs/1104.1607} {\path{arXiv:1104.1607}},
  \href {https://doi.org/10.1016/j.nima.2011.06.103}
  {\path{doi:10.1016/j.nima.2011.06.103}}.

\bibitem{LernerTrigg1991}
R.~G. Lerner, G.~L. Trigg, Encyclopaedia of Physics, 2nd Edition, VHC
  Publishers, 1991.

\bibitem{Chandrasekhar}
S.~Chandrasekhar, Radiative Transfer, Dover Publications, 1960.

\bibitem{Case}
K.~M. Case, P.~F. Zweifel, Linear Transport Theory, Addison-Wesley, New York.

\bibitem{Ishimaru}
A.~Ishimaru, Wave Propagation and Scattering in Random Media, Academic Press,
  New York.

\bibitem{10.1145/2010324.1964951}
E.~D'Eon, G.~Irving, \href{https://doi.org/10.1145/2010324.1964951}{A
  quantized-diffusion model for rendering translucent materials}, ACM Trans.
  Graph. 30~(4) (Jul. 2011).
\newblock \href {https://doi.org/10.1145/2010324.1964951}
  {\path{doi:10.1145/2010324.1964951}}.
\newline\urlprefix\url{https://doi.org/10.1145/2010324.1964951}

\bibitem{SURYAMOHAN20117364}
P.~{Surya Mohan}, T.~Tarvainen, M.~Schweiger, A.~Pulkkinen, S.~R. Arridge,
  \href{http://www.sciencedirect.com/science/article/pii/S0021999111003536}{Variable
  order spherical harmonic expansion scheme for the radiative transport
  equation using finite elements}, Journal of Computational Physics 230~(19)
  (2011) 7364 -- 7383.
\newblock \href {https://doi.org/https://doi.org/10.1016/j.jcp.2011.06.004}
  {\path{doi:https://doi.org/10.1016/j.jcp.2011.06.004}}.
\newline\urlprefix\url{http://www.sciencedirect.com/science/article/pii/S0021999111003536}

\bibitem{Hielscher_1998}
A.~H. Hielscher, R.~E. Alcouffe, R.~L. Barbour,
  \href{https://doi.org/10.1088%2F0031-9155%2F43%2F5%2F017}{Comparison of
  finite-difference transport and diffusion calculations for photon migration
  in homogeneous and heterogeneous tissues}, Physics in Medicine and Biology
  43~(5) (1998) 1285--1302.
\newblock \href {https://doi.org/10.1088/0031-9155/43/5/017}
  {\path{doi:10.1088/0031-9155/43/5/017}}.
\newline\urlprefix\url{https://doi.org/10.1088%2F0031-9155%2F43%2F5%2F017}

\bibitem{GANAPOL2011693}
B.~Ganapol,
  \href{http://www.sciencedirect.com/science/article/pii/S002240731000124X}{Radiative
  transfer with internal reflection via the converged discrete ordinates
  method}, Journal of Quantitative Spectroscopy and Radiative Transfer 112~(4)
  (2011) 693 -- 713, 2009 International Conference on Mathematics and
  Computational Methods (M\&C 2009).
\newblock \href {https://doi.org/https://doi.org/10.1016/j.jqsrt.2010.03.014}
  {\path{doi:https://doi.org/10.1016/j.jqsrt.2010.03.014}}.
\newline\urlprefix\url{http://www.sciencedirect.com/science/article/pii/S002240731000124X}

\bibitem{Wang2023}
C.-H. Wang, X.-Y. Zhang, C.-C. Pan, Z.-Y. Jiang,
  \href{https://link.aps.org/doi/10.1103/PhysRevE.107.045303}{Unified
  discontinuous galerkin finite-element framework for transient conjugated
  radiation-conduction heat transfer}, Phys. Rev. E 107 (2023) 045303.
\newblock \href {https://doi.org/10.1103/PhysRevE.107.045303}
  {\path{doi:10.1103/PhysRevE.107.045303}}.
\newline\urlprefix\url{https://link.aps.org/doi/10.1103/PhysRevE.107.045303}

\bibitem{Ymeli2023a}
G.~L. Ymeli, C.-H. Wang,
  \href{https://link.aps.org/doi/10.1103/PhysRevE.107.015302}{Generalized
  lattice boltzmann method for radiative transfer problem in slab and irregular
  graded-index media}, Phys. Rev. E 107 (2023) 015302.
\newblock \href {https://doi.org/10.1103/PhysRevE.107.015302}
  {\path{doi:10.1103/PhysRevE.107.015302}}.
\newline\urlprefix\url{https://link.aps.org/doi/10.1103/PhysRevE.107.015302}

\bibitem{Feng2021}
L.~Feng, W.~Chen, H.~Liu, On the performance of a mrt lattice boltzmann
  algorithm for transient radiative transfer problems, International
  Communications in Heat and Mass Transfer 128 (2021) 105628.
\newblock \href {https://doi.org/10.1016/j.icheatmasstransfer.2021.105628}
  {\path{doi:10.1016/j.icheatmasstransfer.2021.105628}}.

\bibitem{Ymeli2023b}
J.~Ymeli, A.~Brown, S.~White, Lattice boltzmann method for radiative transfer
  in two-layered slab with graded-index and fresnel reflecting surfaces,
  International Communications in Heat and Mass Transfer 148 (2023) 107025.
\newblock \href {https://doi.org/10.1016/j.icheatmasstransfer.2023.107025}
  {\path{doi:10.1016/j.icheatmasstransfer.2023.107025}}.

\bibitem{Liemert:12}
A.~Liemert, A.~Kienle,
  \href{http://www.osapublishing.org/boe/abstract.cfm?URI=boe-3-3-543}{Infinite
  space green's function of the time-dependent radiative transfer equation},
  Biomed. Opt. Express 3~(3) (2012) 543--551.
\newblock \href {https://doi.org/10.1364/BOE.3.000543}
  {\path{doi:10.1364/BOE.3.000543}}.
\newline\urlprefix\url{http://www.osapublishing.org/boe/abstract.cfm?URI=boe-3-3-543}

\bibitem{PhysRevE.86.036603}
A.~Liemert, A.~Kienle,
  \href{https://link.aps.org/doi/10.1103/PhysRevE.86.036603}{Green's function
  of the time-dependent radiative transport equation in terms of rotated
  spherical harmonics}, Phys. Rev. E 86 (2012) 036603.
\newblock \href {https://doi.org/10.1103/PhysRevE.86.036603}
  {\path{doi:10.1103/PhysRevE.86.036603}}.
\newline\urlprefix\url{https://link.aps.org/doi/10.1103/PhysRevE.86.036603}

\bibitem{ALLAKHVERDIAN2023108726}
V.~Allakhverdian, D.~V. Naumov,
  \href{https://www.sciencedirect.com/science/article/pii/S0022407323002443}{Exact
  analytical solution of the one-dimensional time-dependent radiative transfer
  equation with linear scattering}, Journal of Quantitative Spectroscopy and
  Radiative Transfer 310 (2023) 108726.
\newblock \href {https://doi.org/https://doi.org/10.1016/j.jqsrt.2023.108726}
  {\path{doi:https://doi.org/10.1016/j.jqsrt.2023.108726}}.
\newline\urlprefix\url{https://www.sciencedirect.com/science/article/pii/S0022407323002443}

\bibitem{henyey1941diffuse}
L.~G. Henyey, J.~L. Greenstein, Diffuse radiation in the galaxy, Astrophysical
  Journal 93 (1941) 70--83.

\bibitem{Lepage:1977sw}
G.~P. Lepage, {A New Algorithm for Adaptive Multidimensional Integration}, J.
  Comput. Phys. 27 (1978) 192.
\newblock \href {https://doi.org/10.1016/0021-9991(78)90004-9}
  {\path{doi:10.1016/0021-9991(78)90004-9}}.

\bibitem{Lepage:2020tgj}
G.~P. Lepage, {Adaptive multidimensional integration: VEGAS enhanced}, J.
  Comput. Phys. 439 (2021) 110386.
\newblock \href {http://arxiv.org/abs/2009.05112} {\path{arXiv:2009.05112}},
  \href {https://doi.org/10.1016/j.jcp.2021.110386}
  {\path{doi:10.1016/j.jcp.2021.110386}}.

\end{thebibliography}

\end{document}